%% file: _main_euro.tex
\documentclass[a4paper,fleqn]{cas-dc}

\usepackage[authoryear,longnamesfirst]{natbib}

\def\tsc#1{\csdef{#1}{\textsc{\lowercase{#1}}\xspace}}
\tsc{WGM}
\tsc{QE}

\input{preambule}

\begin{document}
\let\WriteBookmarks\relax
\def\floatpagepagefraction{1}
\def\textpagefraction{.001}

\shorttitle{Survey on Fairness Notions and Related Tensions}    

\shortauthors{G. Alves, F. Bernier, M. Couceiro, K. Makhlouf, C. Palamidessi, S. Zhioua}  

\title[]{Survey on Fairness Notions and Related Tensions}  



%

\author[1]{Guilherme Alves}[orcid=0000-0002-5004-4429]
\cormark[1]
\ead{guilherme.alves-da-silva@loria.fr}
\credit{}

\affiliation[1]{organization={Université de Lorraine, CNRS, Inria N.G.E., LORIA},
            addressline={F-54000}, 
            city={Nancy},
            country={France}}

\author[1]{Fabien Bernier}[orcid=0000-0002-1887-9725]
\ead{fabien.bernier@loria.fr}
\credit{}

\author[1]{Miguel Couceiro}[orcid=0000-0003-2316-7623]
\ead{miguel.couceiro@loria.fr}
\credit{}

\affiliation[2]{organization={Inria, \'Ecole Polytechnique, IPP},
            city={Paris},
            postcode={91120}, 
            country={France}}

\author[2]{Karima Makhlouf}[orcid=0000-0001-6318-0713]
\ead{karima.makhlouf@inria.fr}
\credit{}

\author[2]{Catuscia Palamidessi}[orcid=0000-0003-4597-7002]
\ead{catuscia.palamidessi@inria.fr}
\credit{}

\author[2]{Sami Zhioua}[orcid=0000-0003-2029-175X]
\ead{sami.zhioua@inria.fr}
\credit{}

\cortext[1]{Corresponding author}

\nonumnote{Author ordering on this paper is alphabetical.}
\nonumnote{The research of the first three named authors was partially supported by TAILOR, a project funded by EU Horizon 2020 research and innovation programme under GA No 952215, and the Inria Project Lab ``Hybrid Approaches for Interpretable AI'' (HyAIAI). The research of the last three named authors was supported by HYPATIA, a project funded by the European Research Council (ERC) under the European Union’s Horizon 2020 research and innovation programme under GA No 835294.}


\begin{abstract}
Automated decision systems are increasingly used to take consequential decisions in problems such as job hiring and loan granting with the hope of replacing subjective human decisions with objective machine learning (ML) algorithms. However, ML-based decision systems are prone to bias, which results in yet unfair decisions. Several notions of fairness have been defined in the literature to capture the different subtleties of this ethical and social concept ({\it e.g.,} statistical parity, equal opportunity, etc.). Fairness requirements to be satisfied while learning models created several types of tensions among the different notions of fairness and other desirable properties such as privacy and classification accuracy. \rev{This paper surveys the commonly used fairness notions and discusses the tensions among them with privacy and accuracy. Different methods to address the fairness-accuracy trade-off (classified into four approaches, namely, pre-processing, in-processing, post-processing, and hybrid) are reviewed. The survey is consolidated with experimental analysis carried out on fairness benchmark datasets to illustrate the relationship between fairness measures and accuracy in real-world scenarios.} 
\end{abstract}




\maketitle

\section{Introduction}\label{sec:introduction}

\input{text_euro/1-introduction}

\rev{
\section{Related Work}\label{sec:related_work}
\input{text_euro/2-related_work}

}

\section{Catalogue of fairness notions}\label{sec:notions}
\input{text_euro/3-fairness_notions}

\section{Tensions between fairness notions}\label{sec:notionsTensions}
\input{text_euro/2.5-notionsTensions}

\section{Tensions between fairness and privacy}\label{sec:privacy}
\input{text_euro/3-privacy}

\section{Tension between fairness and classification accuracy}
\label{sec:explanations}
\input{text_euro/4-fairness_accuracy}
\section{Empirical analysis on benchmark datasets}\label{sec:exp}
\input{text_euro/6-experiments}

\section{Conclusion}\label{sec:conclusion}
\input{text_euro/7-conclusion}


\bibliographystyle{cas-model2-names}

\bibliography{orpa_ref,bibFile}

\vfill\eject
\begin{appendix}
\section{Notation Index}\label{a1}
\input{text_euro/8-appendix}
\end{appendix}



\end{document}

%% file: preambule.tex
\usepackage{multirow}
\usepackage{tabularx}
\usepackage{makecell}

\usepackage{graphicx}
\usepackage{subfig}

\usepackage{xspace}
\newcommand{\F}{\textsc{FixOut}\xspace}

\usepackage{colortbl}
\usepackage{xcolor}

\usepackage{lipsum,afterpage,refcount}
\newtheorem{definition}{Definition}
\newtheorem{proposition}{Proposition}
\newtheorem{proof}{Proof}
\newcommand{\setfootnotemark}{%
  \refstepcounter{footnote}%
  \footnotemark[\value{footnote}]}
  
\usepackage[colorinlistoftodos]{todonotes}

\newcommand{\rev}[1]{{\color{black}#1 \color{black}}}
  
\newenvironment{revv}{\par\color{black}}{\par}
 
\usepackage{float}
\usepackage{flafter}

%% file: text_euro/1-introduction.tex

Fairness emerged as an important requirement to guarantee that machine learning (ML) based decision systems can be safely used in practice. Using such systems while fairness is not satisfied can lead to unfair decisions typically discriminating against disadvantaged populations such as racial minorities, women, poverty stricken districts, etc.

With the recent interest for fairness, a multitude of fairness notions have been defined to capture different aspects of fairness. These include statistical group-based notions (e.g., statistical parity~\citep{dwork2012fairness}, equalized odds~\citep{hardt2016equality}, etc.), individual-based notions (e.g., fairness through awareness~\citep{dwork2012fairness}), and causal-based notions (e.g., total effect~\citep{pearl2009causality}, counterfactual fairness~\citep{kusner2017counterfactual}). 
As fairness is a social construct~\citep{jacobs2021measurement} and an ethical concept~\citep{tsamados2022ethics}, defining it is still prone to subjectivity. 
Hence, the aim of replacing the subjective human decisions by objective ML-based decision systems resulted in notions and algorithms still exhibiting unfairness. Hence although the different notions of algorithmic fairness appear internally consistent, several of them cannot hold simultaneously and hence are mutually incompatible~\citep{barocas-hardt-narayanan,mitchell2018prediction,hardt2016equality,berk2021fairness,kleinberg17}. As a consequence, practitioners assessing and/or implementing fairness need to choose among them. 

In addition to the tensions  \rev{between notions} (``intra-notions''), there are tensions between fairness notions and other desirable properties of ML algorithms. One such properties is privacy. This property emerged as a concern that specific information about an individual might be revealed when a model is learned based on a dataset containing that individual. A learning algorithm satisfying privacy will learn aggregate information about the population, but will not incorporate specific information about individuals. Recent results showed that fairness and privacy (differential privacy~\citep{dwork2006differential}) are at odds with each other~\citep{cummings2019compatibility,agarwal21,pujol2020fair}. That is, a learning algorithm that satisfies differential privacy is not guaranteed to generate a classifier that satisfies fairness, unless it has trivial accuracy. 

\rev{
Several methods have been introduced in the literature to reduce bias in the classification output. These methods fall into three categories, namely, pre-processing, in-processing, post-processing. This survey describes a representative set of methods from each category (Reweighting~\citep{calders2009building} and FairBatch~\citep{roh2020fairbatch} for pre-processing, adversarial learning~\citep{zhang2018mitigating}, exponentiated gradient~\citep{agarwal2018reductions}, and AdaFair~\citep{iosifidis2019adafair} for in-processing, and threshold optimizer~\citep{hardt2016equality} and SBD~\citep{fish2016confidence} for post-processing). In addition, we consider a fourth category for methods that combine different fairness interventions (e.g. in-processing and post-processing). We describe a representative set of methods in this category, namely, LFR~\citep{zemel2013learning}, FAE~\citep{iosifidis2019fae}, and FixOut~\citep{alves}.

Reducing bias leads, typically, to a drop in classification accuracy. Therefore, all aforementioned methods can be considered as approaches to tackle the tension between fairness and classification accuracy.


Several surveys of the relatively recent field of ML fairness can be found in the literature~\citep{berk2021fairness,mehrabi2021survey,verma2018fairness,mitchell2018prediction,zliobaite2015survey,gajane2017survey}. However, this survey deviates from existing surveys by focusing on the tensions that exist among fairness notions and between fairness and other desirable ML properties, namely, privacy and classification accuracy. In addition, unlike other surveys, this paper features an experimental analysis based on fairness benchmark datasets. The aim is to show how different unfairness mitigation methods generate classifiers with different fairness but also different accuracy levels.

Section~2 briefly presents commonly used fairness notions spanning all categories (group, individual, and causal) along with their formal definitions. Related work is provided in Section~3. Section~4 describes the tensions and incompatibilities that exist among the various fairness notions. Section~5 shows that fairness and privacy properties are at odds with each other and present a complete formal proof of this incompatibility. Section~6 is a survey on unfairness mitigation methods which trackle the tension between fairness and classification accuracy. 
Section~7 shows how fairness notions can be applied on benchmark and real datasets, and illustrates some of the tensions described in the previous sections. Section~8 concludes and mentions potential directions for future work.
}


%% file: text_euro/2-related_work.tex
With the increasing need for ethical concerns in decision-making systems that have serious implications on individuals and society, several survey papers have been proposed in the literature in the few recent years. In this section, we revisit these survey papers and highlight how our survey deviates from them. 

Makhlouf et al. \citep{DBLP:journals/sigkdd/MakhloufZP21} compiled a survey about existing fairness notions and their main contribution consists of addressing the problem of the applicability of fairness notions to a given real-world scenario. To tackle this problem, the authors identified a set of fairness-related characteristics of the real-world scenario at hand. Then, they analyzed and studied the behavior of each fairness notion. The result of fitting these two elements together consists of a decision diagram
that can be used as a roadmap to guide practitioners and policy makers to select the most suitable fairness notion in a specific setup. However, their survey does not consider conflicts between fairness and privacy. Moreover, our survey provides empirical results that illustrate the tensions between some fairness notions and how the problem of the trade-off between accuracy and fairness can be tackled through explanability and ensemble techniques.

Mehrabi et al.~\citep{mehrabi2021survey} proposed a broader scope for their overview: in addition to 
concisely listing $10$ definitions of fairness metrics, they discussed different sources of bias and different types of discrimination, they listed
methods to mitigate discrimination categorized into pre-processing, in-processing, and post-processing, and they discussed potential directions for contributions in the field. However, they did not discuss any tensions that exist between fairness notions and tensions between fairness, accuracy and privacy which we discuss in depth in this survey.

Arrieta et al.~\citep{arrieta2020explainable} provided an exhaustive overview on explainability in ML. They proposed a novel definition of explainable ML that takes into account the audience for which the explanations are addressed. They also presented a detailed taxonomy of recent contributions related to the explainability of different ML models including explaining Deep Learning methods. Then, they discussed in depth the concept of responsible AI, which imposes the systematic adoption of several AI principles, namely: fairness, accountability and privacy. The survey does not discuss the tensions that might exist between those principles. 

The survey of Mitchell et al. ~\citep{mitchell2018prediction} includes an exhaustive list of group and individual fairness notions and outlines most of the impossibility results among them. They also discussed in detail a ``catalogue'' of choices and assumptions in the context of fairness to address the question of how social goals are formulated into a prediction (ML) problem. Again, their survey does not tackle the problem of tensions between fairness and other ethical considerations (privacy and explainability) in decision making systems as is studied in this paper. 

Tsamados et al.~\citep{tsamados2022ethics} compiled an overview on the ethical problems in AI algorithms and the solutions that have been proposed in the literature. In particular, they provided a conceptual map of six ethical concerns raised by AI algorithms namely: inconclusive, inscrutable, misguided evidence, unfair outcomes, transformative effects, and traceability. The first three concerns refer to epistemic factors, the fourth and the fifth are normative factors and the fifth is relevant both for epistemic and normative factors. The epistemic factors are related to the relevance of the accuracy of the data while the informative factors refer to the ethical impact of AI systems. Although the survey explores a broad scope related to ethical concerns in AI, it remains at a conceptual level and does not address how these ethical concerns are implemented in practice and how they enter in conflict with each other in detail, which we explore in depth in this article.

Other works discussing the trade-off between fairess notions include the work by Kleinberg et al.\citep{kleinberg2016inherent} which discussed the suitability of specific fairness notions in a specific setup. In particular, they discussed the applicability of calibration and balance notions. The survey of Berk et al.~\citep{berk2021fairness} studied the trade-offs between different group fairness notions and between fairness and accuracy in a specific context namely: criminal justice risk assessments. 
They used simple examples based on the confusion matrix to highlight relationships between the fairness notions.

In another research direction, Friedler et al.~\citep{friedler2016possibility} discussed tensions between group fairness and individual fairness. In particular, they defined two worldviews namely: WYSIWYG and WAE. The WYSIWYG (What you see is what you get) worldview assumes that the unobserved (construct) space and observed space are essentially the same while the WAE (we're all equal) worldview implies that there are no inherent differences between groups of individuals based on potential protected attributes. 

A more recent survey by Fioretto et al.~\citep{fioretto2022differential} provided the constraints under which differential privacy and fairness may be achieved simultaneously or having conflicting goals. In particular, they showed that individual fairness and differential privacy can be applied at the same time while group fairness and differential privacy are incompatible. They also examined how and why differential privacy may magnify bias and unfairness in two different settings namely: decision problems and learning tasks. Then, they reviewed existing mitigation measures for the fairness issues arising in these two settings.

%% file: text_euro/3-fairness_notions.tex
Let $V$, $A$, and $X$\footnote{A list of all terms used in this survey appears in Appendix~\ref{a1}
.} be three random variables representing, respectively, the total set of features, the sensitive features, and the remaining features describing an individual such that $V=(X,A)$ and $P(V=v_i)$ represents the probability of drawing an individual with a vector of values $v_i$ from the population. For simplicity, we focus on the case where $A$ is a binary random variable where $A=0$ designates the non-protected group, while $A=1$ designates the protected group. Let $Y$ represent the actual outcome and $\hat{Y}$ represent the outcome returned by the prediction algorithm. Without loss of generality, assume that $Y$ and $\hat{Y}$ are binary random variables where $Y=1$ designates a positive instance, while $Y=0$ a negative one. Typically, the predicted outcome $\hat{Y}$ is derived from a score represented by a random variable $S$ where $\mathbb{P}[S = s]$ is the probability that the score value is equal to $s$.

\rev{
To illustrate the various ML fairness notions, we use a simple job hiring scenario (Table~\ref{tab:example1}). Each sample in the dataset has the following attributes: education level (numerical), job experience (numerical), age (numerical), marital status (categorical), gender (binary) and a label (binary). The sensitive attribute is the applicant's gender, that is, we are focusing on whether male and female applicants are treated equally. Table~\ref{tab:example1}(b) presents the predicted decision  (first column) and the predicted score value (second column) for each sample. The threshold value is set to $0.5$.}

\begin{table*}[!ht]
	\centering
	\caption{\rev{A simple job hiring example. $Y$ represents the data label indicating whether the applicant is hired ($1$) or rejected ($0$). $\hat{Y}$ is the prediction which is based on the score $S$. A threshold of $0.5$ is used.} \medskip}
\label{tab:example1} 
\subfloat{
 \begin{tabular}{cccccc}
    \multicolumn{6}{c}{(a) Dataset} \\
\hline\noalign{\smallskip}
    Gender    & Education Level & Job Experience &  Age  & Marital Status & Y \\
    \noalign{\smallskip}\hline\noalign{\smallskip}
    Female 1           & 8    & 2 &39            & single    & 0 \\
   Female 2            & 8     & 2   & 26             & married     & 1\\
    Female 3            & 12     & 8   & 32             & married     & 1\\
    Female 4     & 11      &  3   & 35             & single     & 0  \\
    Female 5     & 9      &  5   & 29             & married     & 1 \\
    Male 1          & 11    & 3 &34            & single    & 1  \\
    Male 2          & 8     & 0   & 48             & married     & 0 \\
    Male 3        & 7     & 3   & 43             & single     & 1   \\
    Male 4      & 8      &  2   & 26             & married     & 1 \\
    Male 5          &8     & 2   & 41             & single     & 0  \\
    Male 6   & 12      &  8   & 30             & single     & 1  \\
    Male 7   & 10      &  2   & 28             & married     & 1 \\
\noalign{\smallskip}\hline
  \end{tabular}  
  }
 \qquad 
  \subfloat{
   \begin{tabular}{cc}
   \multicolumn{2}{c}{(b) Prediction} \\
\hline\noalign{\smallskip}
    \^{Y} & S\  \\
\noalign{\smallskip}\hline\noalign{\smallskip}
    1 & 0.5\\
    0 & 0.1\\
    1 & 0.5\\
    0 & 0.2\\
    0 & 0.3\\
    1 & 0.8\\
    0 & 0.1\\
    0 & 0.1\\
    1 & 0.5\\
    1 & 0.5\\
    1 & 0.8\\
    0 & 0.3\\
\noalign{\smallskip}\hline
  \end{tabular}
}   
\end{table*}

\paragraph{Statistical parity}~\citep{dwork2012fairness} (a.k.a., \textit{demographic parity} \citep{kusner2017counterfactual}, \textit{independence} \citep{barocas2017fairness}, \textit{equal acceptance rate} \citep{zliobaite2015relation}, \textit{benchmarking} \citep{simoiu2017problem}, \textit{group fairness} \citep{dwork2012fairness}) is one of the most commonly accepted notions of fairness. It requires the prediction to be statistically independent of  the sensitive feature $(\hat{Y}  \perp A)$. 
In other words, the predicted acceptance rates for protected and unprotected groups should be equal. Statistical parity implies that: $$\displaystyle \frac{TP+FP}{TP+FP+FN+TN}\footnote{$TP,FP,FN,$ and $TN$ stand for: true positives, false positives, false negatives, and true negatives, respectively.}$$ is equal for both groups. A classifier \^{Y} satisfies statistical parity if:
\begin{equation}
\label{eq:sp}
\mathbb{P}[\hat{Y} \mid A = 0] = \mathbb{P}[\hat{Y} \mid A = 1].
\end{equation}

\rev{In the example of Table~\ref{tab:example1}, the calculated predicted acceptance rate of hiring male and female applicants is $0.57$ ($4$ out of $7$) and $0.4$ ($2$ out of $5$), respectively. Thus, statistical parity is not satisfied.}

\paragraph{Conditional statistical parity}~\citep{corbett2017algorithmic} (a.k.a., \textit{conditional discrimination-aware classification}~\citep{kamiran2013quantifying}) is a variant of statistical parity obtained by controlling on a set of resolving features (also called explanatory features in \citep{kamiran2013quantifying}). The resolving features (we refer to them as $R$) among $X$ are correlated with the sensitive feature $A$ and give some factual information about the label while leading to a \textit{legitimate} discrimination. Conditional statistical parity holds if:
\begin{equation}
\label{eq:csp}
\begin{split}
\mathbb{P}[\hat{Y}=1 \mid R=r,A = 0] = \mathbb{P}[\hat{Y}=1 \mid R=r,A = 1] \\ \quad \forall r \in range(R).
\end{split}
\end{equation}

\rev{
In the example of Table~\ref{tab:example1}, assuming job experience (denoted $R$) is a resolving variable, there is discrimination against females when $R=2$ or $R=3$ but no discrimination when $R=8$ according to conditional statistical parity.}

\paragraph{Equalized odds}~\citep{hardt2016equality} (a.k.a., \textit{separation}~\citep{barocas2017fairness}, \textit{conditional procedure accuracy equality}~\citep{berk2021fairness}, \textit{disparate mistreatment}~\citep{zafar2017fairness}, \textit{error rate balance}~\citep{chouldechova2017fair}) considers both the predicted and the actual outcomes. The prediction is conditionally independent of the protected feature, given the actual outcome $(\hat{Y} \perp A \mid Y)$. In other words, equalized odds requires that both sub-populations have the same true positive rate $TPR = \frac{TP}{TP+FN}$ and false positive rate $FPR = \frac{FP}{FP+TN}$:
\begin{equation}
\label{eq:eqOdds}
\begin{split}
\mathbb{P}[\hat{Y} = 1 \mid Y=y,\; A=0] = \mathbb{P}[\hat{Y}=1 \mid Y= y,\; A=1] \\  \quad \forall{ y \in \{0,1\}}.
\end{split}
\end{equation}

\rev{
Using the same example (Table~\ref{tab:example1}), the $TPR$ for male and female groups is $0.6$ and $0.33$, respectively, while the $FPR$ is the same ($0.5$) for both groups. Consequently, the equalized odds does not hold, as female candidates are discriminated against.}


Because equalized odds requirement is rarely satisfied in practice, two variants can be obtained by relaxing its equation. The first one is called \textit{equal opportunity}~\citep{hardt2016equality} (a.k.a., \textit{false negative error rate balance}~\citep{chouldechova2017fair}) and is obtained by requiring only TPR equality among groups:
\begin{equation}
\label{eq:eqOpp}
\mathbb{P}[\hat{Y}=1 \mid Y=1,A = 0] = \mathbb{P}[\hat{Y}=1\mid Y=1,A = 1].
\end{equation}

\rev{As $TPR$ does not consider $FP$, equal opportunity is completely insensitive to the number 
of false positives. Equalized odds does not hold in the example of Table~\ref{tab:example1} as $TPR$ is higher for males than females.}

The second relaxed variant of equalized odds is called \textit{predictive equality}~\citep{corbett2017algorithmic} (a.k.a., \textit{false positive error rate balance}~\citep{chouldechova2017fair}) which requires only the FPR to be equal in both groups:
\begin{equation}
\label{eq:predEq}
\mathbb{P}[\hat{Y}=1 \mid Y=0,A = 0] = \mathbb{P}[\hat{Y}=1\mid Y=0,A = 1]. 
\end{equation}
\rev{Since $FPR$ is independent from $FN$, predictive equality is completely insensitive to false negatives. The example in Table~\ref{tab:example1} satisfies predictive parity.}

\paragraph{Conditional use accuracy equality}~\citep{berk2021fairness} (a.k.a., \textit{sufficiency}~\citep{barocas2017fairness}) is achieved when all population groups have equal positive predictive value $PPV=\frac{TP}{TP+FP}$ and negative predictive value $NPV=\frac{TN}{FN+TN}$.  In other words, the probability of subjects with positive predictive value to truly belong to the positive class and the probability of subjects with negative predictive value to truly belong to the negative class should be the same. By contrast to equalized odds, conditional use accuracy equality conditions on the algorithm’s predicted outcome not the actual outcome.  In other words, the emphasis is on the precision of prediction rather than its recall:
\begin{equation} 
\label{eq:condUseAcc}
\begin{split}
\mathbb{P}[Y=y\mid \hat{Y}=y ,A = 0] = \mathbb{P}[Y=y\mid \hat{Y}=y,A = 1] \\ \quad \forall{ y \in \{0,1\}}.
\end{split}
\end{equation}
\rev{
The calculated PPVs for male and female applicants in the hiring example (Table~\ref{tab:example1}) are $0.75$ and $0.5$, respectively. NPVs for male and female applicants are both equal to $0.33$. Overall the dataset in Table~\ref{tab:example1} does not satisfy conditional use accuracy equality. }


\paragraph{Predictive parity}~\citep{chouldechova2017fair} (a.k.a.,  \textit{outcome test} in ~\citep{simoiu2017problem}) is a relaxation of conditional use accuracy equality requiring only equal $PPV$ among groups:
\begin{equation}
\label{eq:predPar}
\mathbb{P}[Y=1 \mid \hat{Y} =1,A = 0] = \mathbb{P}[Y=1\mid \hat{Y} =1,A = 1]  
\end{equation}

\rev{Like predictive equality, predictive parity is insensitive to false negatives. According to predictive parity, there is a discrimination against female candidates in the same example (Table~\ref{tab:example1}).} 

\paragraph{Overall accuracy equality}~\citep{berk2021fairness} is achieved when overall accuracy for both groups is the same. This implies that:  

\begin{equation}
\label{eq:accuracy}
\frac{TP+TN}{TP+FN+FP+TN}
\end{equation}

\noindent is equal for both groups:

\begin{equation}
\label{eq:ovAcc}
\mathbb{P}[\hat{Y} = Y | A = 0] = \mathbb{P}[\hat{Y} = Y | A = 1] 
\end{equation}

Overall accuracy does not hold in Table~\ref{tab:example1} as $\mathbb{P}[\hat{Y} = Y]$ for female candidates is $\frac{2}{5}$ whereas for male candidates, it is $\frac{4}{7}$.

\paragraph{Treatment equality}~\citep{berk2021fairness} is achieved when the ratio of FPs and FNs is the same for both protected and unprotected groups:
\begin{equation}
\label{eq:treatEq}
\frac{FN}{FP} \textsubscript{A=0} = \frac {FN}{FP} \textsubscript{A=1}
\end{equation}

\rev{
For example, in Table~\ref{tab:example1}, the ratio $\frac{FN}{FP}$ for male is $\frac{2}{5}$ whereas for female, it is $\frac{2}{7}$.}

\paragraph{Total fairness}~\citep{berk2021fairness}  holds when all aforementioned fairness notions are satisfied simultaneously, that is, statistical parity, equalized odds, conditional use accuracy equality (hence, overall accuracy equality), and treatment equality. Total fairness is a very strong notion that is very difficult to hold in practice. 

\paragraph{Balance}~\citep{kleinberg17} uses the predicted probability score ($S$) from which the outcome $Y$ is typically derived through thresholding. \textit{Balance for positive class} focuses on the applicants who constitute positive instances and is satisfied if the average score $S$ received by those applicants is the same for both groups:
\begin{equation}
\label{eq:balPosclass}
E[S \mid Y =1,A = 0] = E[S \mid Y =1,A = 1]. 
\end{equation}
\rev{In Table~\ref{tab:example1}, the expected score values are $0.3$ and $0.5$ for hired males and hired females respectively. This also indicates discrimination against females since the latter need a higher score to get hired than males. }
\textit{Balance of negative class} focuses instead on the negative class:
\begin{equation}
\label{eq:balNegclass}
E[S \mid Y =0,A = 0] = E[S \mid Y =0,A = 1].
\end{equation}
\rev{The values in Table~\ref{tab:example1} are $0.35$ and $0.3$ for the non-hired males and females respectively. Hence, there is no balance for negative class, however, there is discrimination \textit{in favor} of female.}


\paragraph{Calibration}~\citep{chouldechova2017fair} (a.k.a. \textit{test-fairness}

\noindent\citep{chouldechova2017fair}, \textit{matching conditional frequencies}

\noindent\citep{hardt2016equality}) holds if, for each predicted probability score $S=s$, individuals in all groups have the same probability to actually belong to the positive class:
\begin{equation}
\label{eq:calib}
\begin{split}
\mathbb{P}[Y =1 \mid S =s,A = 0] = \mathbb{P}[Y =1 \mid S =s,A = 1] \\ \quad \forall s \in [0,1].
\end{split}
\end{equation}

\paragraph{Well-calibration}~\citep{kleinberg17} is a stronger variant of calibration. It requires that (1) calibration is satisfied, (2) the score is interpreted as the probability of truly belonging to the positive class, and (3) for each score $S=s$, the probability of truly belonging to the positive class is equal to that particular score:
\begin{equation}
\label{eq:wellCalib}
\begin{split}
\mathbb{P}[Y =1 \mid S =s,A = 0] = \mathbb{P}[Y =1 \mid S =s,A = 1] = s \\ \quad  \forall \; {s \in [0,1]}.
\end{split}
\end{equation}

Table~\ref{tab:welcal} shows the difference between calibration and well-calibration. The values inside the table represent $\mathbb{P}[Y =y \mid S =s,A = a]$.

\begin{table*}[!h]
	\centering
	\caption{Calibration vs. well-calibration.}
    \label{tab:welcal} 
\subfloat{
\begin{tabular}{lllll}
\multicolumn{5}{c}{(a) Calibrated but not well-calibrated} \\
\hline\noalign{\smallskip}
s & 0.4 & 0.7 & 0.8 & 0.85\\
\noalign{\smallskip}\hline\noalign{\smallskip}
Female        & 0.33 & 0.5 & 0.6 & 0.6\\ 
Male          & 0.33 & 0.5   & 0.6 & 0.6 \\ 
 \noalign{\smallskip}\hline
\end{tabular}
\ }
 \qquad 
 \subfloat{
\begin{tabular}{lllll}
\multicolumn{5}{c}{(b) Calibrated and well-calibrated} \\
\hline\noalign{\smallskip}
s & 0.4 & 0.7 & 0.8 & 0.85\\
\noalign{\smallskip}\hline\noalign{\smallskip}
Female        & 0.4 & 0.7 & 0.8 & 0.85\\ 
Male          & 0.4 & 0.7   & 0.8 & 0.85 \\ 
 \noalign{\smallskip}\hline
\end{tabular}
}   
\end{table*}

\paragraph{Fairness through unawareness} (a.k.a., \textit{blindness, unawareness}~\citep{mitchell2018prediction}, \textit{anti-classification}~\citep{corbett2018measure}, \textit{treatment parity}~\citep{lipton2018does}.) is a simple and straightforward approach to address fairness problem where we ignore completely any sensitive feature while training the prediction system. 

\paragraph{Fairness through awareness}~\citep{dwork2012fairness} (a.k.a., \textit{individual fairness}~\citep{gajane2017survey,kusner2017counterfactual}) implies that similar individuals should have similar predictions. Let $i$ and $j$ be two individuals represented by their attributes values vectors $v_i$ and $v_j$. Let $d(v_i,v_j)$ represent the similarity distance between individuals $i$ and $j$. Let $M(v_i)$ represent the probability distribution over the prediction outcomes. For example, if the outcome is binary ($0$ or $1$), $M(v_i)$ might be $[0.2,0.8]$ which means that for individual $i$, $\mathbb{P}[\hat{Y}=0]) = 0.2$ and $\mathbb{P}[\hat{Y}=1] = 0.8$. Let $d_M$ be a distance metric between probability distributions.  
Fairness through awareness is achieved iff, for any pair of individuals $i$ and $j$: $$d_M(M(v_i), M(v_j))  \leq d(v_i, v_j)$$
In practice, fairness through awareness assumes that the similarity metric is known for each pair of individuals~\citep{kim2018fairness}. A challenging aspect of this approach is the difficulty of determining an appropriate metric function to measure the similarity between two individuals. Typically, this requires careful human intervention from professionals with domain expertise~\citep{kusner2017counterfactual}.

\paragraph{Causality-based fairness notions} differ from all  statistical fairness approaches because they are not totally based on data but consider additional knowledge about the structure of the world in the form of a causal model. 
Therefore, most of these fairness notions are defined in terms of non-observable quantities such as interventions (to simulate random experiments) and counterfactuals (which consider other hypothetical worlds in addition to the actual world).

\begin{figure}[ht]
\centering
\includegraphics [scale=0.27] {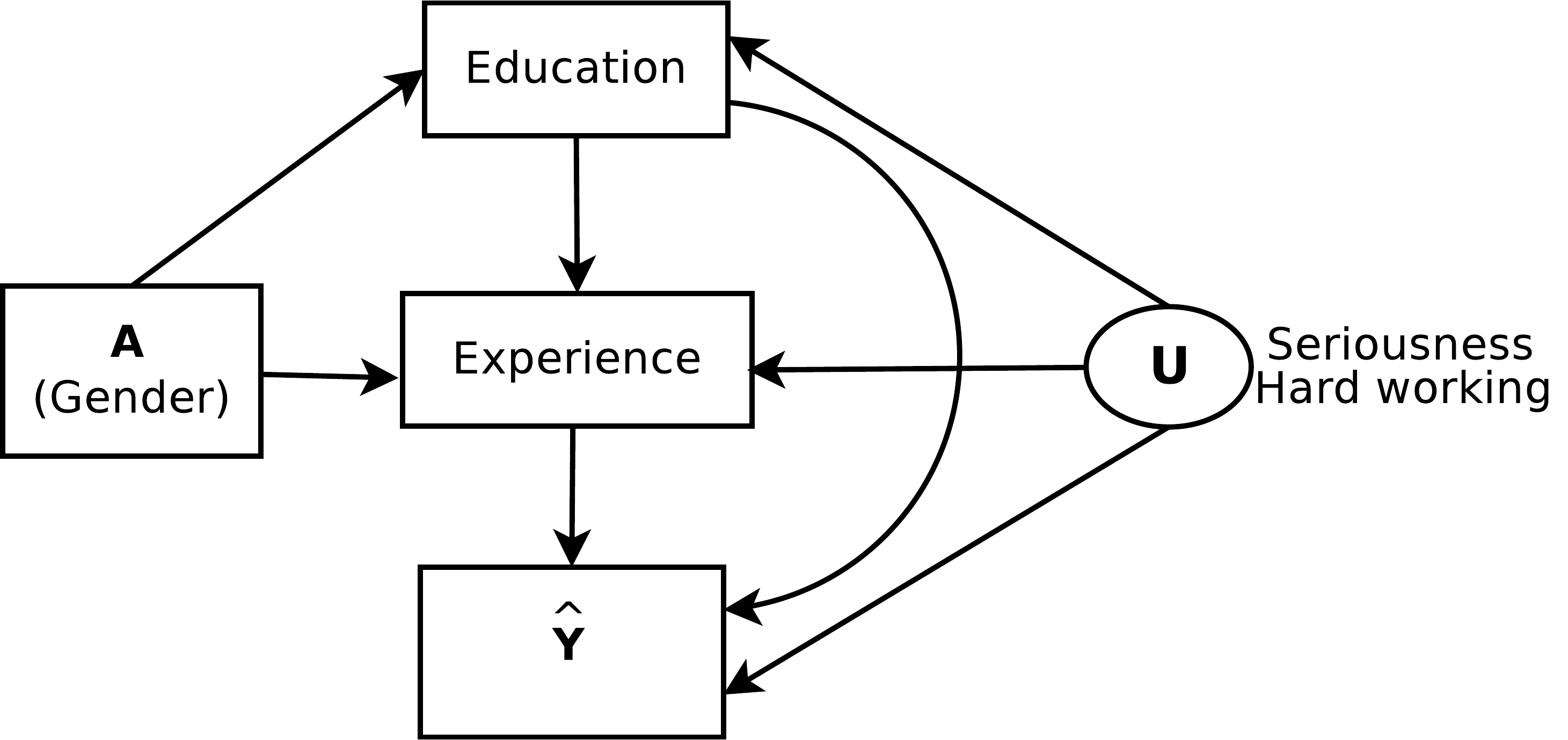}
\caption{A possible causal graph for the hiring example.}
\label{fig:CDCounterfactual}     
\end{figure}

A variable $X$ is a \textit{cause} of a variable $Y$ if $Y$ in any way relies on $X$ for its value~\citep{pearl2016book}. Causal relationships are expressed using structural equations~\citep{bollen1989structural} and represented by causal graphs where nodes represent variables (features), and edges represent causal relationships between variables. Figure~\ref{fig:CDCounterfactual} shows a possible causal graph for the job hiring example where directed edges indicate causal relationships. 

\paragraph{Total effect ($TE$)}~\citep{pearl2009causality} is the causal version of statistical parity and is defined in terms of experimental probabilities as follows: 
\begin{equation}
\label{eq:TE}
TE_{a_1,a_0} (\hat{y}) = \mathbb{P}[\hat{y}_{A\leftarrow a_1}] - \mathbb{P}[\hat{y}_{A\leftarrow a_0}]
\end{equation}
where $\mathbb{P}[\hat{y}_{A\leftarrow a}] = \mathbb{P}[\hat{Y}=\hat{y} \mid do(A=a)]$ is called the experimental probability and is expressed using interventions. An intervention, denoted as $do(V=v)$, is a manipulation of the model that consists in fixing the value of a variable (or a set of variables) to a specific value. Graphically, it consists in discarding all edges incident to the vertex corresponding to variable $V$.  Intuitively, using the job hiring example, while $\mathbb{P}[\hat{Y}=1\mid A=0]$ reflects the probability of hiring among female applicants, $\mathbb{P}[\hat{Y}_{A\leftarrow 0}=1] = \mathbb{P}[\hat{Y}=1 \mid do(A=0)]$ reflects the probability of hiring if \textit{all the candidates in the population} had been female. The obtained distribution $\mathbb{P}[\hat{Y}_{A\leftarrow a}]$ can be considered as a \textit{counterfactual} distribution since the intervention forces $A$ to take a value different from the one it would take in the actual world. Such counterfactual variable is also denoted as $\hat{Y}_{A=a}$ or $\hat{Y}_a$ for short.

$TE$ measures the effect of the change of $A$ from $a_1$ to $a_0$ on $\hat{Y}=\hat{y}$ along all the causal paths from $A$ to $\hat{Y}$. 
Intuitively, while statistical parity reflects the difference in proportions of $\hat{Y}=\hat{y}$ in the current cohort, $TE$ reflects the difference in proportions of $\hat{Y}=\hat{y}$ in the entire population.
A more involved causal-based fairness notion considers the effect of a change in the sensitive feature value (e.g., gender) on the outcome (e.g., probability of hiring), given that we already observed the outcome for that individual. This typically involves an impossible situation that requires returning to the past and changing the sensitive feature value. Mathematically, this can be formalized using counterfactual quantities. The simplest fairness notion using counterfactuals is \textit{the effect of treatment on the treated (ETT)}~\citep{NBERt0107} defined as:
\begin{equation}
\label{eq:ETT}
ETT_{a_1,a_0} (\hat{y}) = \mathbb{P}[\hat{y}_{A\leftarrow a_1} \mid a_0] - \mathbb{P}[\hat{y} \mid a_0]
\end{equation}
$\mathbb{P}[\hat{y}_{A\leftarrow a_1}\mid a_0]$ reads the probability of $\hat{Y}=\hat{y}$ had $A$ been $a_1$, given $A$ had been observed to be $a_0$. For instance, in the job hiring example, $\mathbb{P}[\hat{Y}_{A\leftarrow 1} \mid A=0]$ reads the probability of hiring an applicant had she has been a male, given that the candidate is observed to be female. Such probability involves two worlds: an actual world where $A=a_0$ (the candidate is female) and a counterfactual world where the same individual $A=a_1$ (the same candidate is male). 

\textit{Counterfactual fairness}~\citep{kusner2017counterfactual} is a fine-grained variant of ETT conditioned on all features. That is, a prediction $\hat{Y}$ is counterfactually fair if under any assignment of values $X=x$,
\begin{equation}
\begin{split}
\label{eq:counterfactual}
\mathbb{P}[\hat{Y}_{A\leftarrow a_1}=\hat{y} \mid X=x, A = a_0] = \\ \mathbb{P}[\hat{Y}_{A\leftarrow a_0}=\hat{y} \mid X=x, A = a_0]. 
\end{split}
\end{equation}
It is important to mention that most of causal-based fairness notions require the knowledge of the causal graph. Causal graphs can be set manually by experts in the field, but very often generated using experiments (called also interventions). The process of identifying the causal graph is called causal discovery or structure learning. Binkyte et al.~\citep{binkyte2022causal} show how different causal discovery approaches may result in different causal graphs and, most importantly, how even slight differences between causal graphs can have significant impact on fairness/discrimination conclusions.

\paragraph{Process fairness}
~\citep{grgic2016case} (or \textit{procedural fairness}) can be described as a set of subjective fairness notions that are centered on the process that leads to outcomes. These notions are not focused on the fairness of the outcomes, instead, they quantify the fraction of users that consider fair the use of a particular set of features. 
They are subjective as they depend on user judgments which may be obtained by subjective reasoning.

A natural approach  to improve process fairness is to remove all sensitive (protected or salient) features before training classifiers. This simple approach connects process fairness to fairness through unawareness. 
However, in addition to the proxies problem mentioned at the beginning of Section~\ref{sec:notions}, dropping out sensitive features may impact negatively classification performance~\citep{zafar2017fairness}.

%% file: text_euro/2.5-notionsTensions.tex
It has been proved that there are incompatibilities between fairness notions. For instance, it is not always possible for a predictor to satisfy specific fairness notions simultaneously~\citep{barocas-hardt-narayanan,chouldechova2017fair,zafar2017fairness,mitchell2018prediction}. In the presence of such incompatibilities, the predictor should relax some fairness notions by partially satisfying all of them. 
Incompatibility\footnote{The term impossibility is commonly used as well.} results are well summarized by Mitchell et al.~\citep{mitchell2018prediction} as follows. \rev{Before listing the tensions, it is important to summarize the relationships between fairness notions. In addition, for completeness, we define a new fairness notion, namely, \textit{negative predictive parity}.}

The following proposition formally states the relationship between equalized odds, equal opportunity, and predictive equality.
\begin{proposition}
Satisfying equal opportunity and predictive equality is equivalent to satisfying equalized odds:
$$Eq.~\ref{eq:eqOdds} \Leftrightarrow Eq.~\ref{eq:eqOpp} \wedge Eq.~\ref{eq:predEq}$$
\end{proposition}

Conditional use accuracy equality (Eq.~\ref{eq:condUseAcc}) is ``symmetric'' to equalized odds (Eq.~\ref{eq:eqOdds}) with the only difference of switching $Y$ and $\hat{Y}$. The same holds for equal opportunity (Eq.~\ref{eq:eqOpp}) and predictive parity (Eq.~\ref{eq:predPar}). However, there is no ``symmetric'' notion to predictive equality (Eq.~\ref{eq:predEq}). For completeness, we define such a notion and give it the name  \textit{negative predictive parity}.

\begin{definition}
Negative predictive parity holds iff all sub-groups have the same $NPV = \frac{TN}{FN+TN}$:
\begin{equation}
\label{eq:negpredEq}
P(Y=1 \mid \hat{Y} =0,A = 0) = P(Y=1\mid \hat{Y} =0,A = 1)  
\end{equation}
\end{definition}

\rev{\begin{proposition}
Satisfying equalized odds or conditional use accuracy equality always leads to satisfying overall accuracy. 
$$Eq.~\ref{eq:eqOdds} \vee Eq.~\ref{eq:condUseAcc} \Rightarrow Eq.~(\ref{eq:ovAcc}) $$
\end{proposition}
The reverse, however, is not true. That is, a Machine Learning based Decision Making (MLDM) approach that satisfies overall accuracy does not necessarily satisfy equalized odds or conditional accuracy.}

\paragraph{Statistical parity (independence) \textit{versus} conditional use accuracy equality (sufficiency)}. Independence and sufficiency are incompatible, except when both groups (protected and non-protected) have equal base rates or $\hat{Y}$ and $Y$ are independent. Note, however, that $\hat{Y}$ and $Y$ should not be independent since otherwise, the predictor is completely useless.  More formally,  \\

\begin{tabular}{cccccccc}
\setlength{\tabcolsep}{2pt}
	$\hat{Y}  \perp A$ &  $\wedge$ & $Y \perp A \mid \hat{Y}$  &  $\Longrightarrow$ & \\
	(independence) &  & (strict sufficiency) & & \\
\end{tabular} \\

\begin{tabular}{ccccccccc}
\setlength{\tabcolsep}{2pt}
	& &  $Y \perp A$ & $\vee$ & $\hat{Y} \perp Y$  \\
	& & (equal base rates) & & (useless predictor) & \\
\end{tabular} \\

It is important to mention here that this result does not hold for the relaxation of sufficiency, in particular, predictive parity. Hence, it is possible for the output of a predictor to satisfy statistical parity and predictive parity between two groups having different base rates. 






\paragraph{Statistical parity (independence) \textit{versus} equalized odds (separation)}. Similar to the previous result, independence and separation are mutually exclusive unless base rates are equal or the predictor $\hat{Y}$ is independent of the actual label $Y$~\citep{barocas-hardt-narayanan}. As mentioned earlier, dependence between $\hat{Y}$ and $Y$ is a weak assumption as any useful predictor should satisfy it. More formally, \\

\begin{tabular}{cccccccc}
\setlength{\tabcolsep}{2pt}
	$\hat{Y}  \perp A$ &  $\wedge$ & $\hat{Y} \perp A \mid Y$  &  $\Longrightarrow$ & \\
	(independence) &  & (strict separation) & &  \\
\end{tabular}

\begin{tabular}{cccccccc}
\setlength{\tabcolsep}{2pt}
	& & $Y \perp A$ & $\vee$ & $\hat{Y} \perp Y$ & \\
	& & (equal base rates) & & (useless predictor) & \\
\end{tabular} \\

Considering a relaxation of equalized odds, that is equal opportunity or predictive equality, breaks the incompatibility between independence and separation. 






\paragraph{Equalized odds (separation) \textit{versus} conditional use accuracy equality (sufficiency)}. Separation and sufficiency are mutually exclusive, except in the case where groups have equal base rates. More formally:
\begin{center}
\setlength{\tabcolsep}{0.1pt}
\begin{tabular}{ccccccc}
	$\hat{Y} \perp A \mid Y$ &  $\wedge$ & $Y \perp A \mid \hat{Y}$  & $ \Rightarrow$ &  $Y \perp A$  & & \\
(strict separation) &   & (strict sufficiency) &  &  (equal base rates) &  &  \\
\end{tabular}
\end{center}

Both separation and sufficiency have relaxations. Considering only one relaxation will only drop the incompatibility for extreme and degenerate cases. For example, predictive parity (relaxed version of sufficiency) is still incompatible with separation (equalized odds), except in the following three extreme cases~\citep{chouldechova2017fair}:
\begin{itemize}
\item both groups have equal base rates.
\item both groups have $FPR=0$ and $PPV=1$.
\item both groups have $FPR=0$ and $FNR=1$.
\end{itemize}

The incompatibility disappears completely when considering relaxed versions of both separation and sufficiency.


%% file: text_euro/3-privacy.tex
Privacy in the context of machine learning (ML) is typically formalized using differential privacy~\citep{dwork2006differential}. Differential privacy gives a strong guarantee that the learning algorithm will learn aggregate information about the population but will not encode information about the individuals. Privacy and fairness of ML algorithms have been mainly studied separately. Recently, however, a number of studies focused on the relationship between fairness and privacy~\citep{cummings2019compatibility,pujol2020fair,agarwal21}. These studies attempt to answer two main questions: what is the consequence of guaranteeing fairness on the privacy of individuals? and to which extent the learning accuracy is impacted when fairness and privacy are simultaneously required? It turns out that there is a tension between privacy and fairness. In particular, it is impossible to satisfy exact fairness and differential privacy simultaneously while keeping a useful level of accuracy. Cummings et al.~\citep{cummings2019compatibility} provided a proof of a theorem stating that exact equal opportunity and differential privacy can simultaneously hold only for a constant/trivial classifier (a classifier that outputs always the same decision). \rev{However, the proof contains a flaw. On the other hand, the proof of Agrawal~\citep{agarwal21} does not contain flaws; although it holds onto a relaxed version of fairness, it does not address specifically equal opportunity. 
This section describes complete proof of the impossibility of satisfying simultaneously exact fairness (equal opportunity) and differential privacy while keeping a non-trivial accuracy. Hence, compared to the works of Cummings~\citep{cummings2019compatibility} and Agarwal~\citep{agarwal21}, our proof addresses the flaw of the former and shows explicitly how the proof holds specifically for equal opportunity, which the latter (Agarwal~\citep{agarwal21}) fails to address.}

For the sake of the proof, we use the same variable definitions as in Section~\ref{sec:notions}. In addition, let $\mathcal{X}$ be the data universe consisting of all possible data elements $z=(x,a,y)$ where $x\in X$ are the element's features, $a\in A$ is the sensitive feature, and $y \in Y$ is the actual outcome (label). Let $h:\mathcal{X}\rightarrow \{0,1\}$ be a binary classifier that tries to predict the true outcome $y$ of a data element $z$.
The following definitions are needed for the proof.

\begin{definition}[Trivial classifier]
A classifier $h$ is said to be trivial if it outputs the same outcome independently from the data element inputs:
$$\mathbb{P}[h(z) = \hat{y}] = \mathbb{P}[h(z') = \hat{y}] \qquad \forall z,z'\in \mathcal{X},\quad \hat{y}\in \{0,1\} $$
\end{definition}

\begin{definition}[Datasets adjacency]
A dataset $D$ can be defined in two ways each leading to a different definition of adjacency:
\begin{itemize}
    \item a dataset is a finite set of samples $D=\{z_1,z_2,\ldots,z_n$\} drawn from a distribution over $\mathcal{X}$. With this definition, datasets $D$ and $D'$ are adjacent if they differ in exactly one data element, that is, $z_i \neq z_i'$ for exactly one $i \in [n]$.
    \item a dataset is a distribution over $\mathcal{X}$. With this definition, $D$ and $D'$ are adjacent ($\zeta$-close) if: \begin{equation}\frac{1}{2}\sum_{z\in \mathcal{X}} |\;D(z) - D'(z)\;| \leq \zeta,\label{eq:adjacency}
    \end{equation}
\end{itemize}

where $D(z)$ is the probability of $z$ under distribution $D$.
\end{definition}
\rev{
Equation~\ref{eq:adjacency} can be interpreted as a bounding of the Hamming distance between the two distributions $D$ and $D'$. While the first definition of adjacency focuses on a specific data element $z$, the second definition (Equation~\ref{eq:adjacency}) may involve several data elements $z_1, z_2, \ldots \in \mathcal{X}$.
}

\begin{definition}[Differential privacy]
Let $\mathcal{D}$ be the set of all possible datasets and $\mathcal{R}$ the set of all possible trained classifiers.
A learning algorithm $\mathcal{M}: \mathcal{D} \rightarrow \mathcal{R}$ satisfies $\epsilon$-differential privacy if for any two adjacent datasets $D$, $D' \in \mathcal{D}$, for any $\epsilon < \infty$, and for any subset of models $S \in \mathcal{R}$:
\begin{equation}
\mathbb{P}[\;\mathcal{M}(D) \in S\;] \;\; \leq \;\; e^{\epsilon}\; \mathbb{P}[\;\mathcal{M}(D') \in S\;] \nonumber 
\end{equation}
\end{definition}

$\mathbb{P}[\mathcal{M}(D)\in S]$ represents the probability that a learning algorithm outputs a classifier model in a specific subset of models $S$. 
Hence, to satisfy differential privacy, a learning algorithm should output similar classifiers with similar probabilities on any adjacent datasets.

\begin{proposition}
Every trivial classifier is fair (equal opportunity) and differentially private.
\end{proposition}

\begin{proof}
We first prove that a trivial classifier satisfies always equal opportunity. Then we prove that it always satisfies differential privacy.
Let $h$ be a trivial classifier. Then,
\begin{align}
    \mathbb{P}[h(z) = 1] & = \mathbb{P}[\hat{Y} = 1|Y=y, A=a] \quad \forall z, y, a \\
    & = \mathbb{P}[\hat{Y} = 1|Y=1, A=0] \label{eq:p1-step2}\\
    & = \mathbb{P}[\hat{Y} = 1|Y=1, A=1] \label{eq:p1-step3}
\end{align}
Steps~\ref{eq:p1-step2} and~\ref{eq:p1-step3} correspond to equal opportunity (Equation~\ref{eq:eqOpp}).

For differential privacy, assume that the trivial classifier $h$ outputs $\hat{Y}=1$ with a constant probability $\rho \in ]0,1[$.
Let $D$, $D' \in \mathcal{D}$ be two adjacent datasets. Then,
\begin{align}
\forall z \in d\quad \mathbb{P}[\;h(z)=1\;] & = \rho \\
\forall z' \in d'\quad \mathbb{P}[\;h(z')=1\;] & = \rho
\end{align}
Hence, for any trivial classifier $h$
\begin{align}
\mathbb{P}[\;\mathcal{M}(D) = h\;] = \mathbb{P}[\;\mathcal{M}(D') = h\;] 
\end{align}
\qed
\end{proof}

\begin{proposition}\label{prop2}
No learning algorithm $\mathcal{M}$ can simultaneously satisfy $\epsilon-$differential privacy and guarantee to generate a fair (equal opportunity) classifier which is non-trivial.
\end{proposition}
To prove that Proposition~\ref{prop2} holds, it suffices to find a non-trivial classifier $h$ which is fair on a dataset $D$ and unfair on a neighboring dataset $D'$. This means that $h$ can be generated by a model $\mathcal{M}$ on $D$ but cannot be generated by the same model $\mathcal{M}$ on $D'\in \mathcal{D}$.
\begin{proof}\footnote{The proof is inspired by Cummings et al.~\citep{cummings2019compatibility} and Agarwal~\citep{agarwal21} proofs.}
For any non-trivial classifier $h$, there exist two points $a$ and $b$ such that:
\begin{itemize}
    \item $a$ and $b$ are classified differently ($h(a) \neq h(b)$)\footnote{This is valid for any non-trivial classifier.}
    \item $a$ and $b$ belong to two different groups ($a = (x_1,0,y_1)$ and $b = (x_2,1,y_2)$)\footnote{If $a$ and $b$ belong to the same group, any point in the other group will be different from either $a$ or $b$. So replace $a$ or $b$ with that point.}.
\end{itemize}  
Consider datasets constructed over the following four elements:\\
\begin{center}
$\begin{array}{cc}
    z_1=(x_1,0,1) & \qquad z_2=(x_1,0,0) \\
    z_3=(x_2,1,0) & \qquad z_4=(x_2,1,0)
\end{array}$
\end{center}
Since $h$ is non-trivial and depends only on the observable features ($X$ and $A$), we have: $h(z_1) = h(z_2) = 0$ and $h(z_3) = h(z_4) = 1$.
Let $D$ a dataset over the above four points such that:
\begin{center}
$\begin{array}{cc}
    D(z_1)=\epsilon & \qquad D(z_2)=\frac{1}{2} - \epsilon \\
    D(z_3)=\epsilon & \qquad D(z_4)=\frac{1}{2} - \epsilon
\end{array}$
\end{center}
According to $D$, $h$ is fair for group $A=0$ (most of the points have label $Y=0$ and are all classified $\hat{Y} = 0$) and for group $A=1$ as well (most of the points have label $Y=0$ and are all classified $\hat{Y} = 1$).

Consider now dataset $D'$ on the same four points such that:
\begin{center}
$\begin{array}{cc}
    D'(z_1)=\frac{1}{2} - \epsilon & \qquad D'(z_2)=\epsilon \\
    D'(z_3)=\frac{1}{2} - \epsilon & \qquad D'(z_4)=\epsilon
\end{array}$
\end{center}
According to $D'$, $h$ is (negatively) unfair to group $A=0$ (most of the points have label $Y=1$ but are all classified $\hat{Y} = 0$) and (positively) unfair to group $A=1$ (most of the points have label $Y=0$ but are all classified $\hat{Y} = 1$).
It is important to mention finally that $D$ and $D'$ are not neighbors. However, according to Claim~2 in~\citep{agarwal21}, if a learning algorithm is differentially private, then $\forall D, D'\in \mathcal{D}$, and for all classifiers $h$,
\begin{align}
\mathbb{P}[\;\mathcal{M}(D) = h\;] > 0 &  \Longrightarrow  \mathbb{P}[\;\mathcal{M}(D') = h\;] > 0  
\end{align}
which means that if $h$ can be learned from dataset $D$, it can be also learned from dataset $D'$.

Hence, for any non-trivial classifier $h$ which is fair on a dataset $D$, there always exists another dataset for which $h$ is unfair.
\qed
\end{proof}

%% file: text_euro/4-fairness_accuracy.tex
\begin{revv}
This section focuses on the tension between fairness and classification accuracy, which is also known as the fairness-accuracy trade-off~\citep{kleinberg2016inherent}. 
This tension naturally arises in many real-world scenarios, e.g., mortgage lending~\citep{lee2021algorithmic}.
It is discussed in several papers~\citep{BechavodL17,friedler,menon2018cost,zliobaite2015relation} and it arises once we try to improve fairness in a ML pipeline by using a fairness processor.

Fairness processors are algorithmic approaches (also known as algorithmic interventions and fairness-enhancing interventions) that are conceived to optimize one or more fairness notions.
These approaches are often arranged based on the stage they apply fairness interventions in a ML pipeline: \textit{pre-processing}, \textit{in-processing}, and \textit{post-processing}.
We give an overview of algorithmic interventions in this section. 
Particularly, we focus on methods whose implementations (source codes) are available and that cover all categories.
A more complete list of fairness processors can be found at~\citep{friedler,pessach2022review,orphanou2022mitigating}.

In this survey, we propose the inclusion of a fourth category named \textit{hybrid-processing}, which comprises algorithmic approaches that combine different fairness interventions as a single method and, as a consequence, do not fit in any of the three traditional categories.


\subsection{Pre-processing}
    
    Pre-processing approaches (pre-processors) modify the input in order to achieve fair outcomes. 
    These processors can be applied to any model since they are model-agnostic. 
    However, the fact they change the input before training may harm the explainability. 
    In addition, pre-processors may increase the uncertainty of the classification process which impacts the level of accuracy~\citep{pessach2022review}.
   
    \begin{itemize}
\item \textit{Reweighing}~\citep{calders2009building}. 
This processor assigns different weights to data instances based on the distribution of a sensitive feature and the class label.
The weights are used to guide a sampling procedure (with replacement) in order to obtain a (new) balanced dataset whose sensitive feature and class label are independent.
For instance, data instances that obtained high weights will reappear more often in the balanced dataset.
A classifier is then trained on the balanced dataset.
As a consequence of the sampling procedure, classification errors on the high-weighted data instances are more expensive.

    
   
    
    \item {\textit{FairBatch}}~\citep{roh2020fairbatch}
    This pre-processor is an extension of a batch selection algorithm that modifies the batch training in order to enforce model fairness (e.g., equal opportunity, equalized odds, demographic parity).
    More precisely, it measures fairness and adapts the size of the batch based on sensitive groups (which links this pre-processor with the reweighing approach).

    \end{itemize}
    
\subsection{In-processing}

    In-processing techniques (in-processors) try to change the learning algorithm during the training process.
    Since they are an easy way to impose fairness constraints, these processors usually take into account the tension between fairness and classification performance.
    However, they can not always be applied to any model since they are usually model-specific.
    
    \begin{itemize}
    \item 
    \textit{Debiasing with adversarial learning}~\citep{zhang2018mitigating}. 
    This in-processor trains two neural networks: (1) a predictor and  (2) an adversary.
    Both networks have different objectives since the goal is, at the end of the training process, to attain the separation criterion.
    The goal of the predictor is to learn a function that predicts the class label (or the score in a regression problem),
    while the adversary takes as input the predicted label and its goal is to predict a sensitive feature.
    The predictor has weights $W$ and loss function $L_P$, while the adversary has weights $U$ and loss function $L_A$.
    
    The main idea behind this processor comes in the way the weights of both networks are updated.
    The weights $U$ (of the adversary) are modified based only on the gradient $\nabla_U L_A$. 
    Unlike the adversary's weights, the update of the predictor's weights, $W$, relies on two components: 
    the first one is the gradient $\nabla_W L_P$ (that minimizes predictor's loss function),
    and the second one is $\textup{proj}_{\nabla_W L_A} \nabla_W L_P - \alpha \nabla_W L_A$ that avoids the predictor from helping or not trying to harm the adversary. 
    
    This algorithmic intervention can be applied to classification and regression problems.
    Also, it can improve demographic parity, equalized odds, and equality of opportunity. 
 
    \item
    \textit{Exponentiated gradient} and \textit{grid search reductions} \citep{agarwal2018reductions,agarwal2019fair}. 
These in-processors reduce an unfairness mitigation problem to a sequence of cost-sensitive classification problems.
Here, fairness notions are re-written in a generic constraint format (\textit{i.e.}, a vector of conditional moments), so that processors can support multiple fairness notions.
More precisely, the statistical parity notion is a special case of the generic constraint that follows
\[
    \mu_j(h) = \mathbb{E}[ g_j(X,A,Y,h(X)) \mid \varepsilon_j], 
\]
where $g_j$ is a function that maps a data instance (protected and non-protected features) along with the predicted and actual outcomes into [0,1], $h$ is a classifier, and $\varepsilon_j$ is an event, which is independent of $h$, that relies on features (protected and non-protected) and $Y$.
The idea is then to solve the following problem
\begin{equation}
    M\mu(h) \leq c,
    \label{eq:reduction_pb}
\end{equation}
where $M$ is a matrix $\mathcal{K}\times\mathcal{J}$, and $c$ is a vector describing linear constraints. 
In order to empirically solve Eq.~\ref{eq:reduction_pb}, $\mu(h)$ and $c$ are replaced by $\hat{\mu}(h)$ and $\hat{c}$, respectively, in the form of a Lagrangian 
\[
    L(h,\lambda) = \mathbb{P}[h(X)=Y] + \lambda^T(M\hat{\mu}(h)-\hat{c}),
\]
where $\lambda \in \mathbb{R}_+^{|\mathcal{K}|}$ is a Lagrange multiplier. 
Now, one needs to find a saddle point that is the overlapped point between maximizing $L$ and minimizing $L$. After a few iterations of updating $\lambda$, the optimal $h$ is obtained as a result.

    \item \textit{Prejudice remover regularizer}~\citep{kamishima2012fairness}.
This in-processor relies on the prejudice index (PI) that measures indirect prejudice.
PI quantifies the mutual information between the class label and a sensitive feature, which indicates the degree of dependence on sensitive information.
\[
    PI = \sum_{(Y,A)\in D} \mathbb{P}[Y,A]\ln\left ( \frac{\mathbb{P}[Y,A]}{\mathbb{P}[Y]\mathbb{P}[A]} \right ) 
\]
PI is then included as a regularizer in the optimization function (see below), to take fairness into account.
A parameter $\eta$ is used to (reduce) enforce the importance of (un)fairness in the training process. 
The idea behind the penalty (along with the parameter $\eta$) is to reduce the dependency of the model on sensitive information and its fairness.
\[
    \min_{f} L[f(x),Y] + \eta PI
\]
Prejudice remover can be applied to any discriminative probabilistic classifier.
The original paper employed this in-processor in a logistic regression model.
    
    
    \item {\textit{AdaFair}}~\citep{iosifidis2019adafair}. 
        It is an ensemble approach that trains a set of classifiers (weak learners) and then combines the output of these classifiers as a single outcome.
        More precisely, it is an extension of AdaBoost that takes into consideration Equalized Odds as a fairness notion to be optimized during training.
        In order to do that, it uses the notion of \textit{cumulative fairness} which takes into account the fairness of a weak learner in the past iterations, instead of only considering the current iteration.
        In addition, AdaFair uses confidence scores in the re-weighting process in order to obtain fairer model outcomes, which differs from the traditional AdaBoost which relies only on classification error. 
    \end{itemize}
    
\subsection{Post-processing}
    
    Post-processing approaches (post-processors) modify the algorithm's outputs to satisfy fairness constraints.
    They are usually model-agnostic and are an easier way to improve disparate impact.
    Also, post-processors are agnostic to the input data, which makes them easier to implement.
    However, post-processors usually present low results compared to pre-processors~\citep{kozodoi2022fairness,pessach2022review}.
    
    There exist two categories of post-processors: 
    (1) trans\-parent-box approaches that change the decision boundary of a model, 
    and
    (2) opaque-box approaches that directly change the classification label.
    
    \begin{itemize}
        \item \textit{Threshold Optimizer}~\citep{hardt2016equality}.
        It finds a threshold $\tau$ for each group (w.r.t. a sensitive feature) that minimizes a loss function $L$ and, at the same time, takes into consideration the separation criterion (either equalized odds or equal opportunity).
        This post-processor can be employed on top of any model and it does not require information about non-sensitive features $X$, as it is based only on the joint distribution of the sensitive feature $A$, and the predicted and actual outcomes, $\hat{Y}$ and $Y$ respectively.
        The optimization process searches in the receiver operating characteristic (ROC) curves between true and false positive rates of sub-populations.
        More precisely, for each binary-sensitive feature, this processor finds an intersection point between the ROC curves of the two sub-population, since the shape of the curves and the cost of misclassification are not necessarily the same for different groups.
        In other words, it minimizes the loss of classification while improving equalized odds (or equal opportunity) as described as follows        
\[
\begin{split}
    \min_{\tau}\mathbb{P}[S(X|A = a, Y = 0) \leq  \tau] \cdot L( \hat{Y} = 1, Y = 0) + \\
    (1 - \mathbb{P}[S(X|A = a, Y = 1) > \tau]) \cdot L( \hat{Y} = 0, Y = 1).
\end{split}
\]                
        
        \item \textit{Reject option classification}~\citep{kamiran2012decision}.
        This post-processor requires that classifiers output probability scores.
        Data instances that obtain scores close to 0 or 1 indicate that a classifier has high confidence (low uncertainty) about predicting the class labels of those instances.
        Instead, the Reject option classification focuses on the other group of data instances.
        It relies on the idea that data instances that have high uncertainty can have their class labels switched in order to enforce fairness.
        Once the definition of high uncertainty (critical region) is established by the user, this processor changes the label of instances in the critical region to improve a certain fairness notion.

        \item {\textit{SBD}} (\underline{S}hifted \underline{D}ecision \underline{B}oundary)~\citep{fish2016confidence}.
         This post-processor is also inspired by the boosting mechanism. 
         Unlike the traditional AdaBoost classifier that applies a majority voting to obtain the outcome, SBD uses a confidence score (and not classification error) for aggregating and obtaining the class label.
         Here, the statistical parity notion is incorporated into the confidence score, which is originally defined only by the distance of a data instance to the decision boundary. 
         As a result, this strategy moves the boundary decision towards fairer outcomes (w.r.t. statistical parity).
         SBD can be employed on top of any model but it also allows the use of different fairness metrics.

    \end{itemize}
\end{revv} 

\subsection{Hybrid-processing}
\begin{revv}
Hybrid-processing approaches (hybrid-processors) 
combine more than one algorithmic intervention in an ML pipeline. 
The idea is to use the advantages of a fairness processor to overcome the disadvantages of another processor.     
\end{revv} 

\begin{itemize}

\begin{revv}
\item \textit{LRF} (\underline{L}earning \underline{F}air \underline{R}epresentations)~\citep{zemel2013learning}
This fairness processor transforms (encodes) the data from an original space to a representation space in order to meet the following requirements:
(1) to optimize statistical parity,
(2) to lose the membership information about protected groups while keeping any other information from the original space,
and (3) to map data instances from the representation space to $Y$ such that the mapping has similar performance compared to an optimal classifier, i.e., to keep the highest possible accuracy.
In order to do that, LRF takes into account these requirements by solving an objective function thanks to a 
in-processing approach.
The goal is then to minimize the loss of a multi-objective and non-linear function. 

\item {\textit{FAE}} 
 (\underline{F}airness-\underline{A}ware \underline{E}nsemble)~\citep{iosifidis2019fae} 
is a framework for fairness-aware classification that combines two fairness-enhancing interventions: pre-pro\-cessing and post-processing. 
The first one tackles the problem of group imbalance and class imbalance by generating  samples before the training phase (pre-processing intervention).
It then trains multiple AdaBoost classifiers in order to obtain an ensemble model $E$.
The second one moves the decision boundary of $E$ towards fairer outcomes (post-processing intervention) based on the fairness notion Equal Opportunity. 
FAE sorts the misclassified instances (in descending order) in order to update the parameter $\theta$ that allows it to move the decision boundary.
\end{revv}
\item \textit{\F}
(\underline{F}a\underline{I}rness through e\underline{X}planations and feature drop\underline{Out})~\citep{bhargava,alves2021reducing}
is a framework that produces an ensemble model by combining pre-processing and post-processing interventions. At  pre-processing, \F removes features based on the approach of fairness through unawareness.
This intervention is guided by an explanation method (e.g., LIME~\citep{ribeiro} and SHAP~\citep{shap}) that produces a list $F^{(k)}$ of the $k$ most important features $a_1,a_2, \ldots,a_k$ w.r.t a model $M$ provided by the user.
The framework then applies the following rule to decide whether $M$ is fair: if  $F^{(k)}$ contains sensitive features $a_{j_1},a_{j_2},\ldots ,a_{j_i}$ in $F$ with $i>1$, then $M$ is deemed unfair and the \F's second fairness intervention applies; otherwise, it is considered fair and no action is taken.

In the former case (i.e., $M$ is considered unfair), \F employs {\it feature dropout} and uses the $i$ sensitive features  $a_{j_1},a_{j_2},\ldots ,a_{j_i}\in F^{(k)}$ to build a pool of $i+1$ classifiers in the following way: 
\begin{itemize}
    \item[$\diamond$] for each $1\leq t\leq i$, \F trains a classifier $M_t$  after removing $a_{j_t}$ from $D$, 
    \item[$\diamond$] and an additional classifier $M_{i+1}$ trained after removing all sensitive features $F$ from $D$. 
\end{itemize}
This pool of classifiers is then used to construct an ensemble classifier $M_{final}$. Note that, \F uses only classifiers that provide probabilities.

At the post-processing intervention, \F can use one of these aggregation functions in order to obtain a single outcome and to enforce fairness, namely: \textit{simple}, \textit{weighted}, and \textit{learned weighted} averages.

In the simple average, all outputs have the same importance, even though some classifiers might be fairer than others. 
Given a data instance $x$ and a class $C$, for an ensemble classifier $M_{final}$ that uses simple averaging, the probability of $x$ being in class $C$ is computed as follows

\begin{equation} \label{eq:simple_avg}
    P_{M_{final}}(x \in C) =  \frac{1}{i+1}\sum_{t=1}^{i+1} P_{M_t}(x \in C),
\end{equation}

\noindent where $P_{M_t}(x \in C)$ is the probability predicted by model $M_t$. 

\F can improve fairness w.r.t. multiple sensitive features. However, it relies on explanations that can be vulnerable to adversarial attacks in some  contexts.\\

\end{itemize}

\begin{revv}

\begin{table*}[]
\footnotesize
\setlength{\tabcolsep}{5pt}
\renewcommand{\arraystretch}{2}
\caption{\rev{Summary of some fairness-enhancing interventions.}}
\begin{tabular}{|l|l|l|l|l|}
\hline
\rotatebox[origin=c]{90}{\textbf{Type}} & \textbf{Method} & \textbf{Fairness notion}  & \textbf{Type of model}   & \makecell[l]{\textbf{Artifact} \\ \textbf{availability}} \\ \hline
\multirow{3}{*}{\rotatebox[origin=c]{90}{Pre-proc.}} & Reweighing~\citep{calders2009building} & Statistical parity  & Model agnostic  & AIF 360 \\ \cline{2-5}
& FairBatch~\citep{roh2020fairbatch}   & \makecell[l]{Equalized odds,   \\ Statistical parity} & Model agnostic  & - \\ \cline{2-5}
& \makecell[l]{Debiasing  with \\ adversarial learning~\citep{zhang2018mitigating}}    &  Separation criterion  &  Gradient-based  & AIF 360  \\ \hline
\multirow{3}{*}{\rotatebox[origin=c]{90}{In-proc.}} 
& \makecell[l]{Exponentiated \\ Gradient~\citep{agarwal2018reductions,agarwal2019fair}}   &  \makecell[l]{Equalized odds, \\ Statistical parity}  & Model agnostic & \makecell[l]{AIF 360, \\ Fairlearn} \\ \cline{2-5}
& Prejudice  Remover~\citep{kamishima2012fairness} & \makecell[l]{Normalized \\ prejudice index}  & Logistic regression  & AIF 360 \\ \cline{2-5}
& AdaFair~\citep{iosifidis2019adafair}    & Equalized odds & AdaBoost  & Git repository\setfootnotemark\label{2}         \\ \hline
\multirow{3}{*}{\rotatebox[origin=c]{90}{Post-proc.}} 
& \makecell[l]{Threshold \\ Optimizer~\citep{hardt2016equality}}       & Equalized odds    & Any score based & \makecell[l]{AIF 360, \\ Fairlearn}   \\ \cline{2-5}
& \makecell[l]{Reject option \\ classification~\citep{kamiran2012decision}}            &   Independence criterion  & Model agnostic  &  AIF 360 \\ \cline{2-5}
& SBD~\citep{fish2016confidence}        & Statistical parity&  Any score based &  Git repository\setfootnotemark\label{sbd}        \\ \hline 
\multirow{3}{*}{\rotatebox[origin=c]{90}{Hybrid}}
& \makecell[l]{LRF~\citep{zemel2013learning}} &  Statistical parity &  Logistic regression &  AIF 360        \\ \cline{2-5}
& FAE~\citep{iosifidis2019fae}        & Equal opportunity & \makecell[l]{Bagging and \\ boosting based}  & Git repository\setfootnotemark\label{3}         \\ \cline{2-5}
& FixOut~\citep{bhargava,alves2021reducing}     & Process fairness  & Any score based & Git repository\setfootnotemark\label{4}  \\ \hline       
\end{tabular}
\afterpage{
\footnotetext[\getrefnumber{2}]{\url{https://github.com/iosifidisvasileios/AdaFair}}
\footnotetext[\getrefnumber{sbd}]{\url{https://github.com/j2kun/fkl-SDM16}}
\footnotetext[\getrefnumber{3}]{\url{https://github.com/iosifidisvasileios/Fairness-Aware-Ensemble-Framework}}
\footnotetext[\getrefnumber{4}]{\url{https://gitlab.inria.fr/galvesda/fixout}}
}
\label{tb:fair-proc}
\end{table*}

Table~\ref{tb:fair-proc} presents a summary of the fairness-enhancing interventions described here. 
It also indicates the links for the code artifacts (git repositories or the Python packages for fairness: AIF 360\footnote{\url{https://github.com/Trusted-AI/AIF360}} or Fairlearn\footnote{\url{https://github.com/fairlearn/fairlearn}}).


\subsection{Assumptions and expectations for the fairness-accuracy trade-off}

Even though various fairness processors take into account both fairness and classification accuracy during the fairness intervention, there is still room for studying, characterizing, and defining this trade-off.
On the one hand, distinct conclusions have been found about the impact on classification accuracy when fairness is enforced. 
For instance, one can say that improving fairness can compromise accuracy~\citep{kleinberg2016inherent}, however, in specific contexts, it can actually increase accuracy~\citep{pessach2021improving,wick2019unlocking}.

On the other hand, other papers have been focused on characterizing or questioning the underlying assumptions made in previously published studies.
For instance,~\citep{zliobaite2015relation} shows that in the evaluation of the fairness-accuracy trade-off, the acceptance rate must be taken into account since classification accuracy from distinct acceptance rates can not be comparable. More precisely, they rely on a notion of discrimination to assess fairness and show that better classification accuracy does not necessarily mean better classification if it comes from distinct acceptance rates.
More recently,~\citep{cooper2021emergent} argue that  researchers make assumptions that may lead to actually unfairness outcomes (or emergent unfairness).
More precisely, three unsuitable assumptions are indicated: 
(1) fairness metrics are sufficient to assess fairness, 
(2) the lack of consideration of historical context, 
and (3) collecting more data on protected groups as an adequate solution.

\end{revv}

%% file: text_euro/6-experiments.tex
To show how fairness notions are used to assess fairness and to illustrate some of the tensions described above, three benchmark datasets are used, namely, \textit{communities and crimes, German credit,} and \textit{Compas}. 
For each one of them, the most common fairness notions are computed in four scenarios: baseline model (logistic regression including all the features in the dataset), \rev{baseline model after Reweighing (pre-processor), 
baseline model along with Threshold Optimizer (post-processor),
and hybrid-processor\footnote{
This hybrid-processor is equivalent to the unfairness mitigation step of \F. 
} 
 (of pre-processing along with post-processing)} using logistic regression. 
This allows us to highlight tensions between fairness notions and to show how feature dropping through process fairness produces an ensemble classifier with a good trade-off between fairness and classification accuracy.

\subsection{Communities and crimes}
\label{subsec:com_crimes}
The \textit{communities and crimes} dataset\footnote{https://archive.ics.uci.edu/ml/datasets/communities+and+crime} includes information relevant to per capita violent crime rates in several communities in the United States and the goal is to predict this crime rate. The dataset includes a total number of $123$ numerical features and $1994$ instances. $22$ features have been dropped as they contain more than $80\%$ missing values. The label \textit{violent crime rate} has been transformed into a binary feature by thresholding\footnote{The mean value of the violent crime rate in the dataset is used as a threshold.} where $1$ corresponds to high violent rate and $0$ corresponds to low violent rate. To assess fairness, we consider two different settings depending on the sensitive feature at hand. First, the \textit{communities racial makeup} is considered as the sensitive feature thus, two groups are created, namely: whites (communities with a high rate of whites) and non-whites (communities with a high rate of non-whites\footnote{Blacks, Asians, Indians, and Hispanics are grouped into a single group called non-whites}). Second, the \textit{communities rate of divorced female} is used as a sensitive feature where we divide the samples into two sub-populations based on whether the rate of divorced females in a community is high ($1$) or low ($0$)\footnote{The mean value of the divorced female rate in the dataset is used as a threshold.}.

\begin{figure}[!ht]
    \centering
    \includegraphics[scale=0.35]{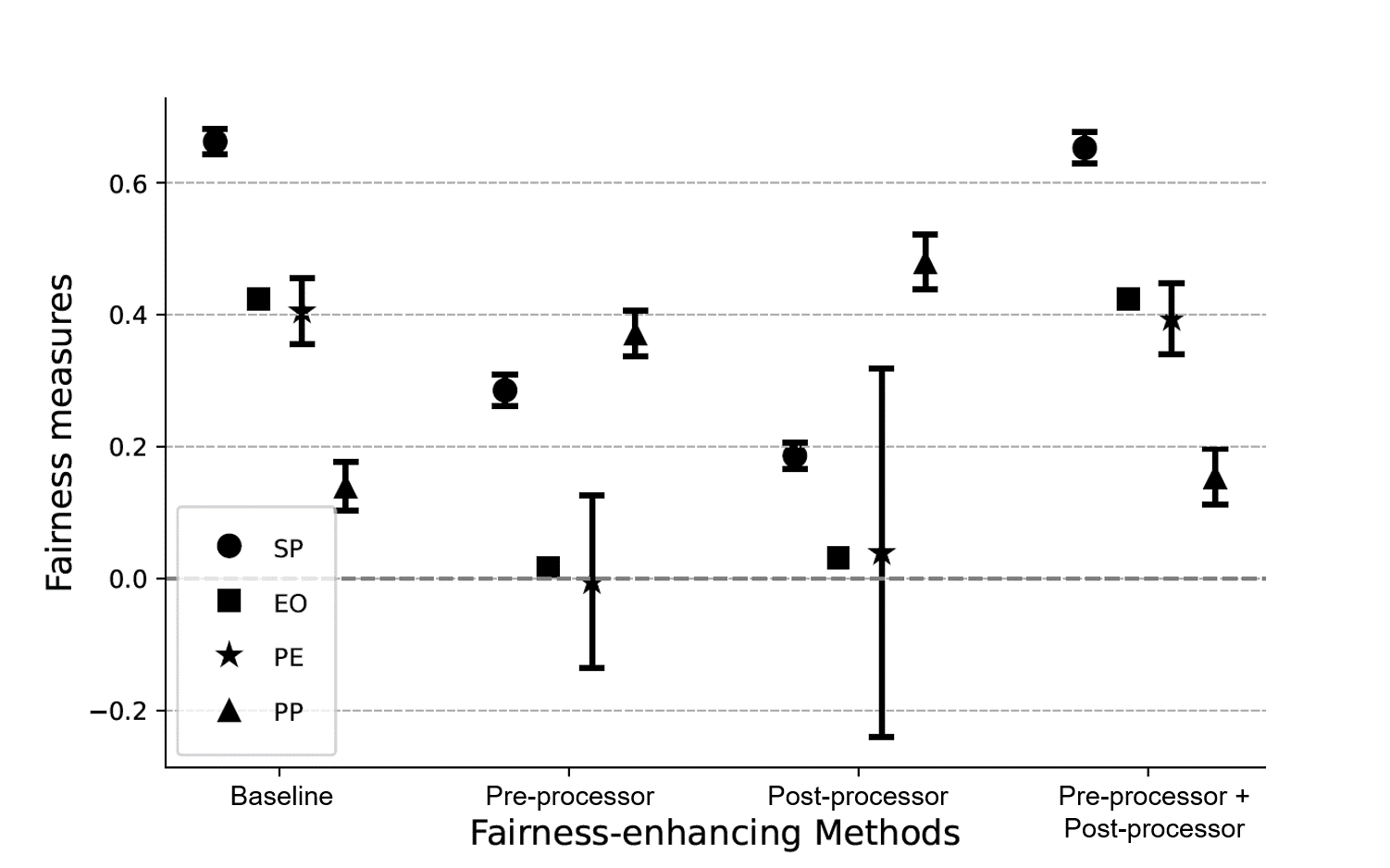} 
    \caption{Fairness assessment for the communities and crimes dataset with race as a sensitive feature.}
    \label{fig:comm_raceResults}
\end{figure}

\begin{figure}[!ht]
    \centering
    \includegraphics[scale=0.35]{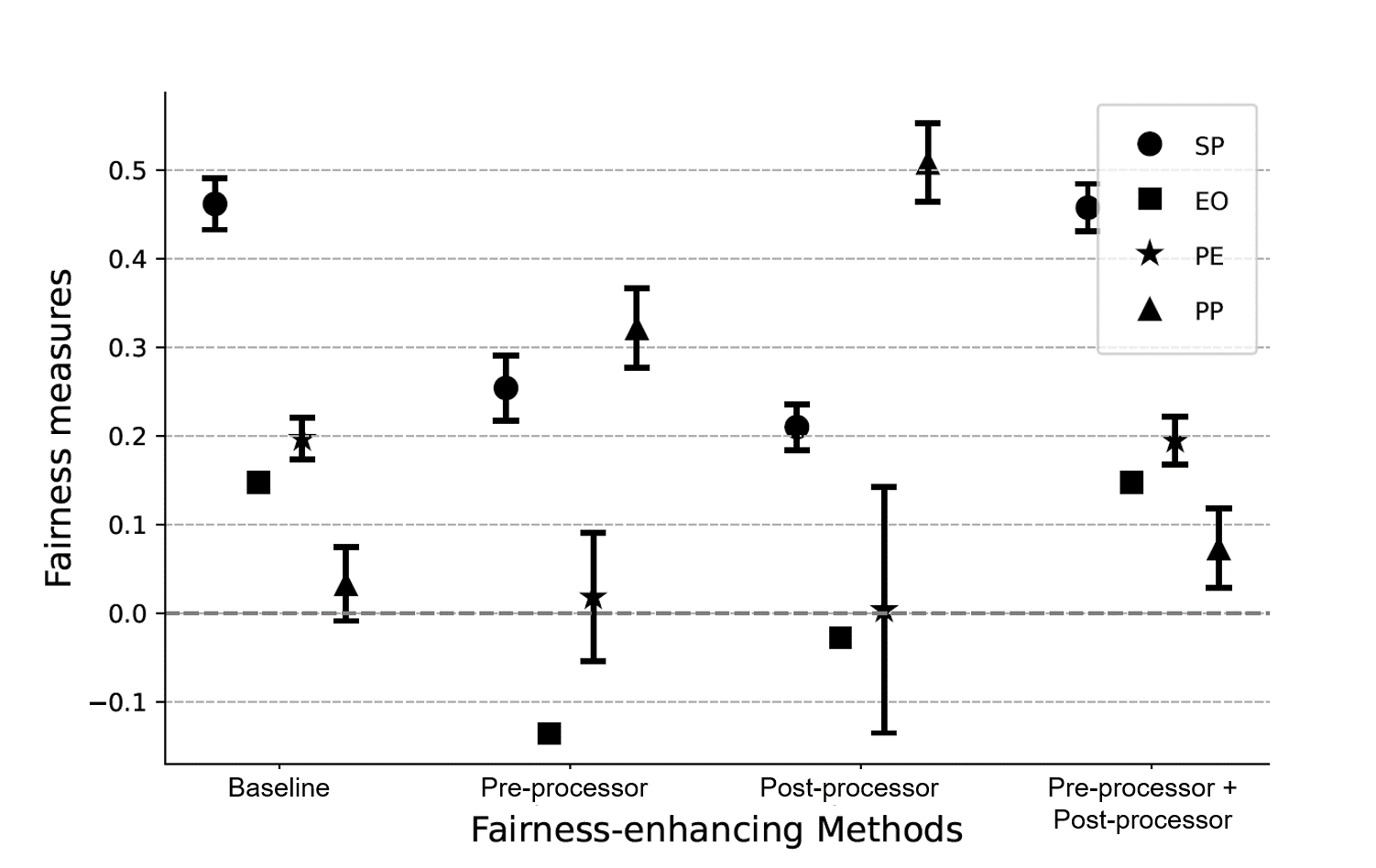} 
    \caption{Fairness assessment for the communities and crimes dataset with divorced female rate as a sensitive feature.}
    \label{fig:comm_divResults}
\end{figure}

\begin{revv}
Figures~\ref{fig:comm_raceResults} and~\ref{fig:comm_divResults} show fairness assessment results for the \textit{communities and crimes} dataset using the baseline model and then the hybrid-processor. For both models, we applied the ten-fold cross-validation technique, using $90\%$ of the data for training and the remaining $10\%$ of the data for testing. Five fairness notions are applied, namely: statistical parity (SP), equal opportunity (EO), predictive equality (PE), and predictive parity (PP). 
The results show discrimination against communities with a high rate of non-whites in the first setting and against communities with a high rate of divorced females in the second setting for all fairness notions.
Fairness processors decrease the difference in some cases, in particular, EO was improved by the pre-processor and the post-processor, while the hybrid-processor reduces the discrimination w.r.t. to PP values in the first setting (Fig.~\ref{fig:comm_raceResults}). In the second setting (Fig.~\ref{fig:comm_divResults}), the pre-processor and the post-processor have inverted the discrimination w.r.t. EO.
\end{revv}


Process fairness empirical analysis focuses 
on the impact of feature dropout on classifiers' dependence on sensitive features.
The results are shown in Table~\ref{tab:proFair_communitiesResults}.
Column ``Contribution'' contains the average value of feature contribution throughout the cross-validation protocol.
Column ``Ranking'' presents the average position of features in the top $k$ most important features;
here, we adopted $k=20$ for all experiments.
We can observe that (the absolute value of) contributions of both sensitive features decrease when we use the hybrid-processor (pre-processing and post-processing), e.g., the absolute value of the contribution of ``Divorced female rate'' decreases from $0.0199$ (baseline) to $0.0080$ (hybrid).
By analyzing the ranking, one notes that the position of both sensitive features decreases, i.e., the position in the list of most import features moves down, which indicates that they become less important compared to other features (ranking positions increase).
For instance, ``Race'' moved from 7.9 (baseline) to 15.5 position (hybrid), i.e., it is closer to the end of the list. \rev{Classification accuracy for the hybrid approach, however, remains exactly the same as the baseline case (see Figure~\ref{fig:accuracy}) while the accuracy of the other two fairness processors dropped.}

\begin{table}[htpb]
\centering
\caption{Process fairness assessment for the \textit{communities and crimes} dataset.}
\label{tab:proFair_communitiesResults}
\setlength{\tabcolsep}{2pt}
\small
\begin{tabular}{l|ll|ll}
\hline
          & \multicolumn{2}{c}{Contribution} & \multicolumn{2}{c}{Ranking}  \\
          & Baseline   & Hybrid   & Baseline & Hybrid \\ 
\hline
Race                 & 0.0092     & 0.0027              & 7.9      & 15.5 \\ 
Divorced female rate & -0.0199    & -0.0080             & 1.6      & 6.5  \\ 
\hline
\end{tabular}
\end{table}

\subsection{German credit}
\label{subsec:german}
The \textit{German credit} dataset\footnote{https://archive-beta.ics.uci.edu/ml/datasets/statlog+german+credit+data} is composed of the data of $1000$ individuals applying for loans. Among $21$ features in the dataset, $7$ are numerical and $13$ are categorical. Numerical and binary features are used directly as features in the classification and each categorical feature is transformed to a set of binary features, arriving at $27$ features in total. This dataset is designed for binary classification to predict whether an individual will default on the loan ($1$) or not ($0$). We consider first, \textit{gender} as a sensitive feature where female applicants are compared to male applicants. Then, \textit{age} is treated as a protected feature where the population is divided into two groups based on whether they are above or below the mean age in the dataset ($35.5$ years old). 

\begin{figure}[!ht]
    \centering
    \includegraphics[scale=0.35]{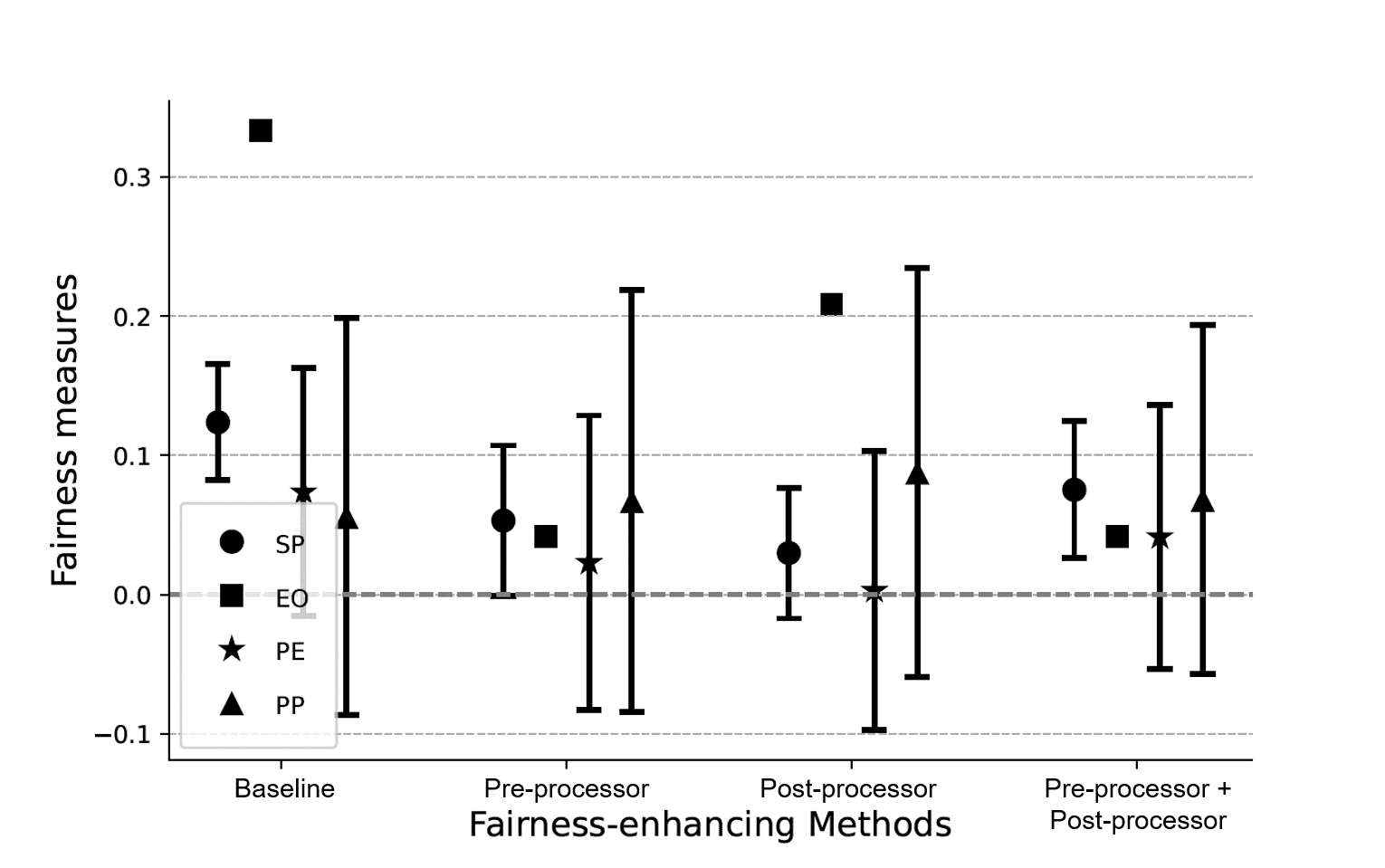} 
    \caption{Fairness assessment for the German credit dataset with sex as a sensitive feature}
    \label{fig:german_sexResults}
\end{figure}

\begin{figure}[!ht]
    \centering
    \includegraphics[scale=0.35]{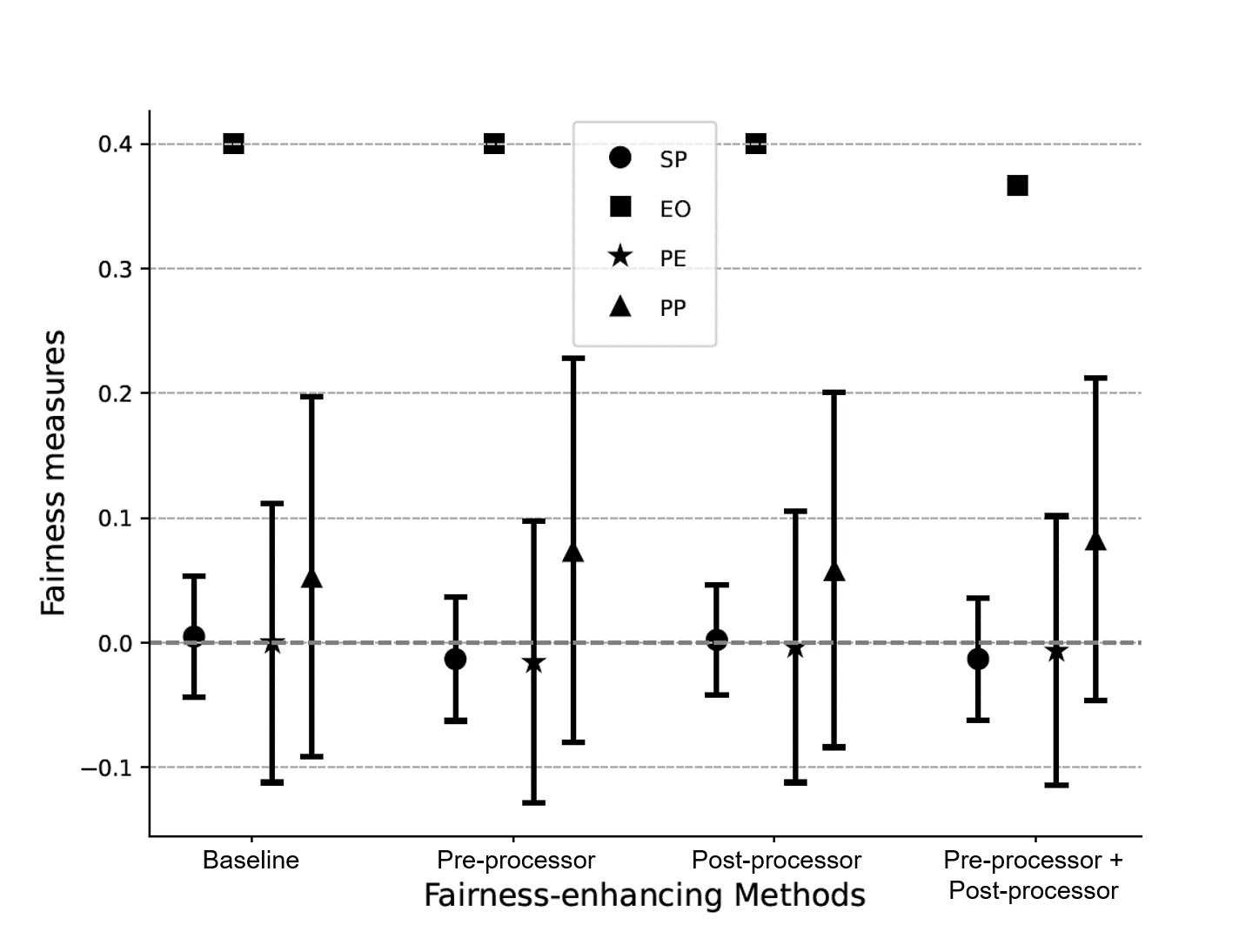} 
    \caption{Fairness assessment for the German credit dataset with age as a sensitive feature}
    \label{fig:german_ageResults}
\end{figure}

\rev{Figures~\ref{fig:german_sexResults} and~\ref{fig:german_ageResults}  show the results for assessing fairness notions for the \textit{German credit} dataset. As for the communities and crimes dataset, four models are trained using $10$-fold cross-validation, namely, baseline (logistic regression), baseline with a pre-processor, baseline with a post-processor, and the hybrid-processor (combination of pre-processor and post-processor). 
Overall, results for all methods show that the applicants who are above the mean age are discriminated against compared to the applicants under the mean age based on EO and PP  regardless of the sensitive feature used (gender and age). 
That is, male and older applicants are privileged over female and younger applicants, respectively, when applying EO and PP.
However, in the second setting (age as a sensitive feature, Fig.~\ref{fig:german_ageResults}) there is parity when SP is used to assess fairness (SP close to 0) while the disparity of PP  increases. }
Divergence between SP and PP is an example of the first incompatibility result in Section~\ref{sec:notionsTensions}. 


Table~\ref{tab:proFairgermanResults} shows the contribution of the sensitive feature on the classification output for the baseline as well as the hybrid approach. Notice that the configuration is the same as the \textit{communities and crime} case.
Similarly to~\textit{communities and crime}, the hybrid-processor improves the contribution and ranking of ``Age'' compared to the baseline.
However, 
the hybrid-processor only improved the ranking of ``Gender'' but not the contribution of this feature. Classification accuracy has slightly dropped from $0.71$ in the baseline model to $0.69$ in the hybrid-processor.

\begin{table}[htpb]
\centering
\caption{Process fairness assessment for the German credit dataset.}
\label{tab:proFairgermanResults}
\begin{tabular}{l|ll|ll}
\hline
          & \multicolumn{2}{c}{Contribution} & \multicolumn{2}{c}{Ranking} \\ 
          & Baseline   & Hybrid   & Baseline & Hybrid \\ 
\hline
Age     & -0.0111 & -0.0060 & 11.0 & 14.1 \\ 
Gender  & -0.0001 & 0.0020  & 15.0 & 17.6 \\ 
\hline
\end{tabular}
\end{table}

\subsection{Compas}
\label{subsec:compas}
The \textit{Compas} dataset contains information from Broward County, Florida, initially compiled by ProPublica~\citep{angwin2016machine} and the goal is to predict the two-year violent recidivism. That is, whether a convicted individual would commit a violent crime in the following two years ($1$) or not ($0$). Only black and white defendants who were assigned \textit{Compas} risk scores within $30$ days of their arrest are kept for analysis~\citep{angwin2016machine} leading to $5915$ individuals in total. We consider \textit{race} as a sensitive feature in the first setting and \textit{gender} in the second. Each categorical feature is transformed into a set of binary features leading to $11$ features in total. 

\begin{figure}[!ht]
    \centering
    \includegraphics[scale=0.35]{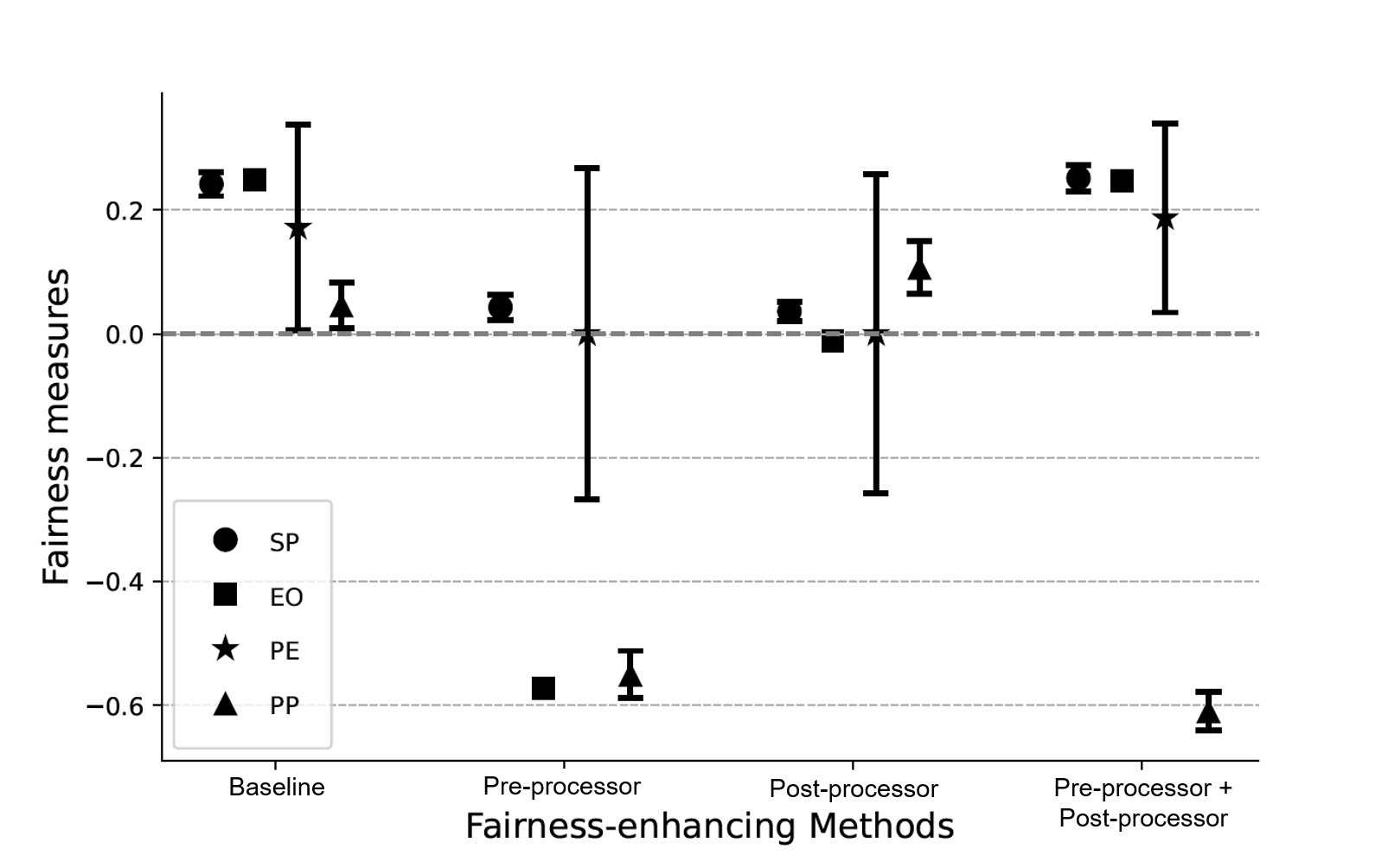} 
    \caption{Fairness assessment for the Compas dataset with race as a sensitive feature.}
    \label{fig:compas_raceResults}
\end{figure}

\begin{figure}[!ht]
    \centering
    \includegraphics[scale=0.35]{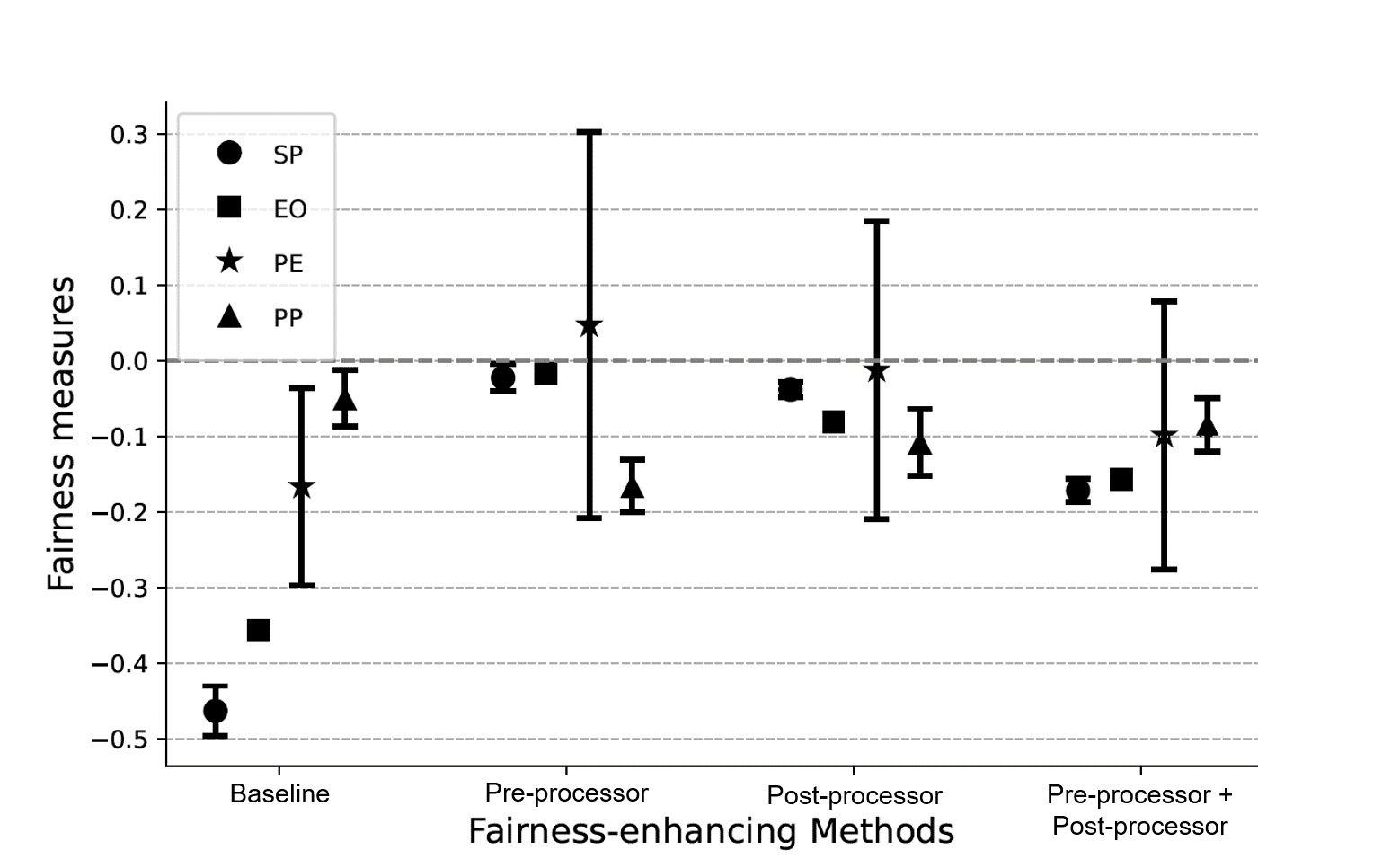} 
    \caption{Fairness assessment for the Compas dataset with sex as a sensitive feature.}
    \label{fig:compas_sexResults}
\end{figure}

Similarly to the previous experiments, \rev{Figures~\ref{fig:compas_raceResults} and~\ref{fig:compas_sexResults} show the four fairness notions results for the baseline and the three fairness-enhancing methods (pre-processor, post-processor, and the hybrid combination of pre-processor and post-processor). 
The figures show similar findings as those discussed in the \textit{German credit} use case. That is, SP, EO, and PE are not satisfied for both settings (blacks vs. whites and females vs. males) while we have at least one fairness notion closer to 0 (e.g. PP without any fairness intervention and PE after applying either the pre-processor or the post-processor).}
This corroborates the debate that has arisen between Propublica and Northpointe\footnote{Now Equivant.} (\textit{Compas} designers) where Propublica used EO and PE to prove that \textit{Compas} privileges whites over blacks. On the other hand, the Northepointe's answer was that PP is a more suitable fairness notion to apply and they proved that \textit{Compas} satisfies PP for blacks and whites~\citep{dieterich2016compas}. 


\begin{figure}[!ht]
    \centering
    \includegraphics[scale=0.35]{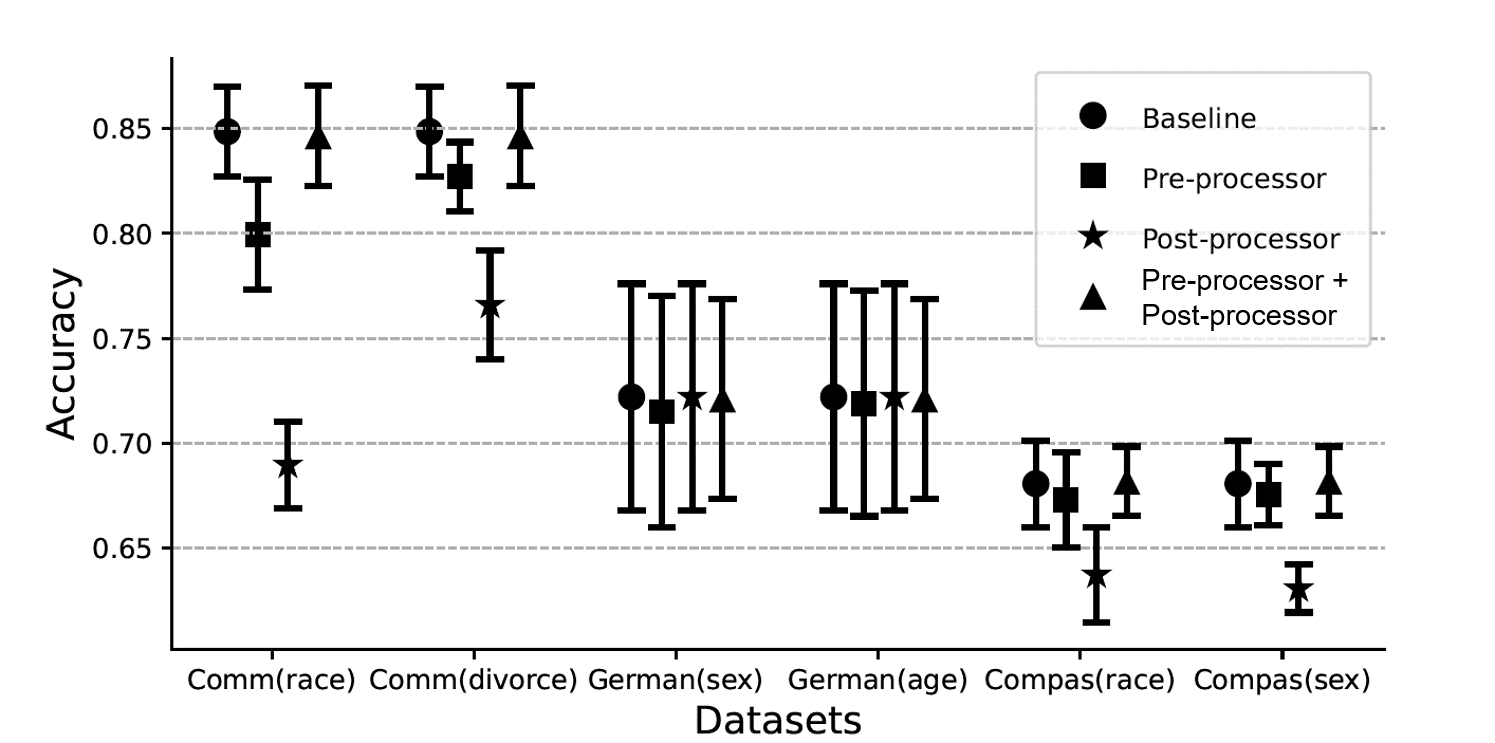} 
    \caption{Accuracy of all datasets after applying the different fairness-enhancing methods.}
    \label{fig:accuracy}
\end{figure}

For process fairness (Table~\ref{tab:proFaircompasResults}), similarly to the previous benchmark datasets, the contribution of ``Race'' decreases when using the hybrid-processor.
In the same way, the ranking of this feature increases from $7.1$ in the case of the baseline model to $8.5$ in the case of the hybrid-processor.
Surprisingly, LIME explanations did not report ``Gender'' as a highly important feature for baseline outcomes;
this feature was already in the last position in the ranking (with no contribution). 
As a result, we do not see any decrease w.r.t feature contribution and ranking. \rev{Note finally that classification accuracy is almost the same for the hybrid-processor, the original model, and the pre-processor, but it decreased when the post-processor was applied (see Figure~\ref{fig:accuracy}).}

\begin{table}[htpb]
\centering
\caption{Process fairness assessment for the Compas dataset.}
\label{tab:proFaircompasResults}
\begin{tabular}{l|ll|ll}
\hline
          & \multicolumn{2}{c}{Contribution} & \multicolumn{2}{c}{Ranking} \\ 
          & Baseline   & Hybrid   & Baseline & Hybrid \\ 
\hline
Race & -0.0017 & -0.0003 & 7.7  & 8.5 \\ 
Gender  & 0.0000  & 0.0000  & 10.0 & 10.0 \\ 
\hline
\end{tabular}
\end{table}

%% file: text_euro/7-conclusion.tex
Implementing fairness is essential to guarantee that ML-based automated decision systems produce unbiased decisions and hence avoid unintentional discrimination against some sub-populations (typically minorities). This survey discusses two important issues related to implementing fairness. 

First, there are several reasonable fairness requirements that can be satisfied simultaneously. This means that fairness practitioners have to choose among them. Second, implementing fairness can create tensions with other desirable properties of ML algorithms, such as privacy and classification accuracy. 
\rev{Empirical results showed that among the unfairness mitigation methods considered in this survey, pre-processing (reweighting) and post-processing (threshold optimizer) are the most efficient to mitigate bias. However, the hybrid approach produced the best accuracy among all unfairness mitigation methods. This survey highlights the need to construct fair ML algorithms that address appropriately the different types of tensions.


 The most recent fairness notions in the literature are causal-based~\citep{makhlouf2020survey} and reflect the now widely accepted idea that using causality is necessary to appropriately address the problem of fairness. Hence, a promising future work is to study how the tensions described in this survey are reflected in causal-based fairness notions. For instance, enforcing causal-based fairness notions relax the tension with privacy and/or accuracy or amplify them. Besides, the most recent fairness notions, however, are causal-based and reflect the now widely accepted idea that using causality is necessary to appropriately address the problem of fairness.
}

%% file: text_euro/8-appendix.tex
\begin{tabular}{ll}
\hline
$V$ & set of attributes  \\
$A$ & sensitive attributes  \\
$X$ & remaining (non-sensitive) attributes \\
$Y$ & actual outcome \\
$\hat{Y}$ &  outcome returned\\
$S$ & score \\
$R$ & resolving features \\
$M$ & pre-trained classifier \\
$D$ & dataset \\
$F$ & list of features contributions \\
$F^{(k)}$ & list of the $k$ most important features  \\
$E$ & explanation method \\
$x$ & data instance\\
$f(x)$ & outcome of a classifier  \\
$g$ & linear (interpretable) model  \\
$z$ & interpretable representation of $x$ \\
$h_x(z)$ & transformation function \\
$K$ & maximum coalition size \\
$\pi$ & kernel (LIME,SHAP) \\
$\sigma$ & kernel-width \\
$\Omega$ & measure of complexity\\
$d$ & distance function  \\
$\mathcal{B}$ & desired number of explanations \\
$\mathcal{V}$ & selected instances \\
$W$ & explanation matrix \\
$I$ & array of feature importance \\
$w_t$ & weight of the $t$-th classifier \\
$C$ & class (label) \\
$a_{i}$ & $i$-th attribute (feature) \\
$c_{i}$ & global feature contribution associated with $a_{i}$ \\
\hline
\end{tabular}